\documentclass[onecolumn]{emulateapj}

\usepackage{natbib}

\newcommand{\dfrac}[2]{{\displaystyle \frac{#1}{#2}}  }

\newcommand{\eqref}[1]{(\ref{#1})}

\def\lesssim{\mathrel{\hbox{\rlap{\hbox{\lower4pt\hbox{$\sim$}}}\hbox{$<$}}}}
\def\gtrsim{\mathrel{\hbox{\rlap{\hbox{\lower4pt\hbox{$\sim$}}}\hbox{$>$}}}}

\slugcomment{ApJ accepted}

\shorttitle{Disk-Planet Interaction and Dynamical Friction}
\shortauthors{Muto, Takeuchi, Ida}

\begin{document}

\title{On the Interaction between a Protoplanetary Disk and a Planet in
an Eccentric Orbit: Application of Dynamical Friction}

\author{Takayuki Muto\altaffilmark{1}}
\author{Taku Takeuchi}
\author{Shigeru Ida}
\affil{Department of Earth and Planetary Sciences, 
Tokyo Institute of Technology, \\
2-12-1 Oh-okayama, Meguro-ku, Tokyo, 152-8551, Japan}

\email{muto@geo.titech.ac.jp}
\altaffiltext{1}{JSPS Research Fellow}

\begin{abstract}
 We present a new analytic approach to the disk-planet interaction that
 is especially useful for planets with eccentricity larger than the disk
 aspect ratio.  We make use of the dynamical friction
 formula to calculate the force exerted on the planet by the disk, and
 the force is averaged over the period of the planet.  The resulting
 migration and eccentricity damping timescale agrees very well with the
 previous works in which the planet eccentricity is  
 moderately larger than the disk aspect ratio.  
 The advantage of this approach is that it is possible to
 apply this formulation to arbitrary large eccentricity.  We have found
 that the timescale of the orbital evolution depends largely on the
 adopted disk model in the case of highly eccentric planets.  
 We discuss the possible implication of our results to the theory of
 planet formation. 
\end{abstract}


\section{Introduction} 
\label{sec:intro}

The gravitational interaction between a planet and a protoplanetary disk
is one of the main topics of the theory of planet formation.  
A low-mass planet embedded in a protoplanetary disk interacts with the
disk by the gravitational force, and its orbital elements change as a
result of the interaction.  The change of the semimajor axis of the
planet is called (type I) orbital migration.  
It has recently been noted that the direction of migration is sensitive
to the disk model if the planet is in a circular orbit
\citep[e.g.,][]{PM06,PBCK10,PBK10}.  

The observations of extrasolar planets have revealed that there are a
number of planets with high eccentricity, and the median eccentricity is
$\sim 0.3$ \citep{US07}.  
One interesting question here is whether it is possible to have 
an eccentric planet in a circular disk.  
Recent numerical simulations \citep{CDKN07,BK10} show that the
eccentricity always damps.  
It is reported that the eccentricity damping and the 
migration timescales do not strongly depend on the physical state of
the gas (e.g., radiative or locally isothermal) for a planet with high
eccentricity, and the timescale becomes longer if the eccentricity
becomes larger.  

The linear analysis of the interaction between the disk and a
planet has been done by a number of authors
\citep[e.g.,][]{GT80,Art93,TW04} for a planet with low eccentricity.
In particular, \citet{PL00} obtained the eccentricity damping
timescale and migration timescale for a planet with $e\gtrsim h$, 
where $e$ is the eccentricity of the planet and $h$ is the disk aspect
ratio ($h=H/r$, where $H$ is the disk scale height and $r$ is the
disk radius), but their approach is restricted to $e \ll 1$.  
\citet{TW04} performed three-dimensional modified local linear
analyses 
\footnote{Here, ``modified'' means that they take the terms up to the
second lowest order of $H/r$.}
to calculate the gravitational interaction between a disk and a planet
in an eccentric orbit.  They have calculated the density perturbation at
every location of the orbit, and therefore, it is possible to calculate
the instantaneous force acting on the planet at every position on the
orbit.  
The instantaneous force acting on the planet is then averaged over the
orbital period to obtain the timescales of the orbital evolution.  We
note that since they use the (modified) local approximations,
their results can be applied to the case where the eccentricity of the
planet is small compared to the disk aspect ratio.

In this paper, we present an analytical model for the interaction
between a low-mass planet in an eccentric orbit and the disk.  
In contrast to the previous approach in which one uses Fourier 
decomposition and calculates the contributions 
from Lindblad and corotation resonances \citep[e.g.,][]{PL00}, we 
make use of the dynamical friction formula to estimate the force acting
on the planet at every location on the orbit.  
The force is then averaged over the orbital period to
obtain the evolution timescales of the orbital parameters.  
Using this ``real-space'' model, 
it is also possible to obtain more intuitive pictures 
of disk-planet interaction.  
We note that most of the highly eccentric planets detected so far are
gas giants, but we consider a low-mass planet in this paper to make the 
problem simpler.  Our analytic approach would reveal an underlying
physical mechanism of the planet-disk interaction, which, we believe,
would be useful in the investigation of high-mass planets as well.  
We also note that in the course of the formation processes of the
planets, it is important to understand the evolution of the orbital
parameters of a low-mass protoplanet or the core of the gas giants.  In
this case, our approach based on linear perturbation analyses may be
directly applicable.

Our approach is similar to that of \citet{TW04} in the sense that we
calculate the force acting on the planet at every instance of the orbit.
The major difference, however, is that we use a very simple model for
the force acting on the planet.
As we shall show later in this paper, our approach is 
especially useful for the case with eccentricity larger than the disk
aspect ratio, and we expect that it is possible to use our model for a
planet with an arbitrary large eccentricity. 
Therefore, we are in the parameter space that is complementary to
\citet{TW04}.  
However, it is noted that the use of a simple model also poses the
limitation of our approach, and we shall discuss the applicability of
the model towards the end of the paper.

The dynamical friction in a gaseous medium itself is also an interesting 
topic investigated by a number of authors.  
\citet{RS80} investigated the stationary
pattern around a gravitating object in a homogeneous medium using a
linear perturbation analysis, and 
concluded that the dynamical friction force vanishes if the
particle's speed is subsonic, while the dynamical friction force 
varies as $v^{-2}$ in the supersonic case, where $v$ is the speed of the
particle.  
The result in the supersonic case is in agreement with the case of the
collisionless system \citep{Ch43}.   

The conclusion of zero dynamical friction force seems rather
counterintuitive since the drag force may 
experience the sudden drop when the particle's speed becomes from
supersonic to subsonic.
This apparent contradiction 
is resolved by \citet{Ost99}, who performed the 
time-dependent analysis.  She showed that the dynamical friction
force depends linearly with the speed of the particle when it moves at a
subsonic velocity, while the force depends on $v^{-2}$ for the supersonic
case.  The results of \citet{Ost99} obtained by linear perturbation
analyses are in good agreement with non-linear numerical simulations
\citep{SSB99}.   

After these works, there are a number of works on this topic with
different configurations of the problem 
(e.g., \citet{KEK05,KK07,KK09}), but
it seems there has not been a work considering a slab geometry, which
can be applied to the problem of disk-planet interaction.  

In this paper, we first perform a linear perturbation analyses and
derive an analytical formula of 
the dynamical friction force exerted on a particle embedden in a
homogeneous slab of gas.   
We then apply the formula to the problem of the disk-planet interaction.
We consider the case where the planet is in an eccentric orbit.  
Although the gas flow in a protoplanetary disk is not homogeneous, we
show that it is actually possible to obtain reasonable results if the
planet is in an eccentric orbit.   
We note that 
the results of the dynamical friction force may have different
astrophysical applications from the problem of the disk-planet
interaction. 

This paper is constructed as follows.  
In Section \ref{sec:df}, we derive the dynamical friction force exerted
on a point particle embedded in a homogeneous gas slab.  
In Section \ref{sec:diskplanet}, we apply the
dynamical friction formula obtained in Section \ref{sec:df} to the
problem of disk-planet interaction.  We then discuss some possible
applications to the planet formation theory in Section 
\ref{sec:discuss}, and Section \ref{sec:summary} is for summary.

\section{Dynamical Friction in a Gaseous Slab}
\label{sec:df}

In this section, we consider the dynamical interaction between a point
particle and a homogeneous, viscous gaseous slab 
in which the particle is embedded.  
We show that it is possible to obtain a dynamical
friction formula whose behavior is similar to that of \citet{Ost99} in
this setup, 
although there is additional dependence on the viscosity as well as the 
Mach number of the particle.  We also show that in the limit of a high
Mach number, the resulting formula does not depend on the viscosity.

\subsection{Basic Setup}

We consider a homogeneous gaseous slab with surface density $\Sigma_0$.
We take the Cartesian coordinate system $(x,y)$ in the plane of the
slab, and consider a point particle with mass $M$ traveling in the
$y$-direction at a constant speed $v_0$.  
The basic equations we consider
are the (vertically integrated) hydrodynamic equations 
with a simplified prescription of viscosity, 
\begin{equation}
 \dfrac{\partial \Sigma}{\partial t} + \nabla \cdot
  \left( \Sigma \mathbf{v} \right) = 0,
  \label{EoC_full}
\end{equation}
\begin{equation}
 \dfrac{\partial \mathbf{v}}{\partial t} 
  + \mathbf{v}\cdot \nabla \mathbf{v} 
  = -\frac{1}{\Sigma}\nabla{P} + \nu \nabla^2 \mathbf{v}
  - \nabla \psi,
  \label{EoM_full}
\end{equation}
where $\Sigma$ is the surface density, $\mathbf{v}$ is the velocity, $P$
is the (vertically integrated) pressure, 
$\nu$ is the viscous coefficient, and $\psi$ is the
gravitational potential of the particle.  We use a simple form of
viscosity in order to keep the problem simple and tractable. 
We use a simple, isothermal equation of state
\begin{equation}
 P = c^2 \Sigma,
\end{equation}
where $c$ is the sound speed.  We consider an inertial coordinate system
in which the particle is always at the origin so the background flow 
velocity is given by $\mathbf{v}=v_0 \mathbf{e}_y$.  For the
gravitational potential, we make use of the form that is commonly
incorporated in the investigation of the disk-planet interaction in
two-dimensional analyses,
\begin{equation}
 \psi = - \dfrac{GM}{\sqrt{x^2 + y^2 + \epsilon^2}},
  \label{pot_2D}
\end{equation}
where $\epsilon$ is the softening length, which is a sizable
fraction of the thickness of the slab in order to mimic the
three-dimensional effects.  

This form of the gravitational potential \eqref{pot_2D} 
is the model for the vertically averaged gravitational potential, and
therefore, the softening length should be of the order of the vertical
scale length of the disk.  
It is to be noted that, with this form of the
potential, the gravitational potential close to the point mass 
(typically, the distance closer than $\epsilon$) is underestimated.  If
the vertical averaging is taken, the potential close to the mass behaves
as $\propto \log r$, where $r$ is the distance from the point mass, in
contrast to equation \eqref{pot_2D}, which 
behaves as constant for $r \lesssim \epsilon$.  

However, as we shall show later, the main contribution to the dynamical
friction force comes from the region $r \sim \epsilon$, and the
contribution from the region $r < \epsilon$ would affect the results by
only some factor.  It is also noted that the dependence of the force on
the physical parameters can be captured with the form of the potential
given by equation \eqref{pot_2D}.  
Therefore, we use this form of the potential as a model of the 
vertically-averaged potential in this paper.  
The outcome of this approximation will be discussed in detail later in
this section.  
We also note that 
this form of the potential is widely used in the numerical calculations
of disk-planet interaction.  
One benefit of using the simple form of the gravitational potential
given by equation \eqref{pot_2D} is that it is possible to obtain an
analytic expression for the dynamical friction.

\subsection{Linear Perturbation Analysis}
\label{dynamicalfric_lin}

We now perform the linear perturbation analysis to calculate the surface
density perturbation induced by the particle's gravity.  
We use the subscript zero to indicate the background quantities and 
we denote all the
perturbed quantities by $\delta$, e.g., surface density is given by
$\Sigma=\Sigma_0 + \delta \Sigma$.  
We retain the terms up to the first order of the perturbation.  
Equations \eqref{EoC_full} and \eqref{EoM_full} now become
\begin{equation}
 \dfrac{\partial}{\partial t} \dfrac{\delta \Sigma}{\Sigma_0}
  + v_0 \dfrac{\partial}{\partial y} \dfrac{\delta \Sigma}{\Sigma_0} 
  + \dfrac{\partial}{\partial x} \delta v_x 
  + \dfrac{\partial}{\partial y} \delta v_y = 0,
  \label{EoC_lin}
\end{equation} 
\begin{equation}
 \left[ \dfrac{\partial}{\partial t} 
  + v_0 \dfrac{\partial}{\partial y} - \nu \nabla^2
  \right] \delta v_x = -c^2 \dfrac{\partial}{\partial x}
 \dfrac{\delta \Sigma}{\Sigma_0} 
 - \dfrac{\partial \psi}{\partial x},
 \label{EoMx_lin}
\end{equation}
\begin{equation}
 \left[ \dfrac{\partial}{\partial t} 
  + v_0 \dfrac{\partial}{\partial y} - \nu \nabla^2
  \right] \delta v_y = -c^2 \dfrac{\partial}{\partial y}
 \dfrac{\delta \Sigma}{\Sigma_0} 
 - \dfrac{\partial \psi}{\partial y}.
 \label{EoMy_lin}
\end{equation}
From these equations, we derive a single second-order differential
equation for the surface density perturbation,
\begin{equation}
 \left( \dfrac{\partial}{\partial t} 
  + v_0 \dfrac{\partial}{\partial y} \right)^2
 \alpha
 - \nu \nabla^2 
 \left( \dfrac{\partial}{\partial t} 
  + v_0 \dfrac{\partial}{\partial y} \right)
 \alpha
 - c^2 \nabla^2 \alpha
 = \nabla^2 \psi,
 \label{wave_lin}
\end{equation}
where we have defined $\alpha \equiv \delta \Sigma/\Sigma_0$ 
and $\nabla^2 = (\partial/\partial x)^2 + (\partial/\partial y)^2$
is the Laplacian.

We now consider a steady state where $\partial/\partial t=0$.  
In \citet{Ost99}, it is pointed out that it is necessary to perform a
time-dependent analysis in order to correctly obtain the dynamical
friction force, especially when the particle's velocity is subsonic.  
This is because the contribution to the force coming from the place very
far away from the particle is not negligible.  
However, as we shall show below, the analysis assuming the steady
state is adequate in the slab geometry we consider in this section
since the perturbation induced at a distant place from 
the particle does not contribute to the force.  

In Appendix \ref{app:timedep}, 
we explicitly show that the force coming from the
time-dependent terms falls as $t^{-1}$.  
Here, we briefly show this by the order-of-magnitude estimate.  
If we consider the place far away from the particle, the gravitational
potential is given by 
\begin{equation}
 \psi \sim \dfrac{GM}{r},
\end{equation}
where $r$ is the distance from the particle.  
We expect that this gravitational energy is of the same order of the
magnitude with the perturbed thermal energy of the gas.  In the
three-dimensional analysis, we therefore expect that 
\begin{equation}
 \dfrac{\delta \rho}{\rho_0} \sim \dfrac{GM}{c^2 r},
\end{equation}  
where $\rho$ is the gas density.
In the slab geometry, we expect that
\begin{equation}
 \dfrac{\delta \Sigma}{\Sigma_0} \sim \dfrac{GM}{c^2 r},
\end{equation}  
where $\Sigma$ is the surface density.
In the three-dimensional analysis, the force $\delta F$ 
acting on the particle from
the gas shell at the distance $r$ with the width $\delta r$ is
\begin{equation}
 \delta F \sim \dfrac{GM \delta \rho r^2 \delta r}{r^2}
  \sim \frac{(GM)^2 \rho \delta r}{c^2 r} \propto \dfrac{\delta r}{r}.
  \label{estimate_3D}
\end{equation} 
The force from the shell decays only with $r^{-1}$. 
Therefore, when summed over all the shells, the contribution from the
distant shell is not negligible.
\footnote{This is why Coulomb logarithm is involved in the dynamical
friction formula.}
In the case of two-dimensional slab geometry, on the other hand, the
force from the ring at distance $r$ with width $\delta r$ is
\begin{equation}
 \delta F \sim \dfrac{GM \delta \Sigma r \delta r}{r^2}
  \sim \dfrac{(GM)^2 \Sigma_0 \delta r}{c^2 r^2} 
  \propto \dfrac{\delta r}{r^2}.
  \label{estimate_2D}
\end{equation}
The force from the distant ring decays as $r^{-2}$ and therefore, the
contribution from the distant ring does not account the total force when
summed over all the rings.  We note here that the contribution to the
dynamical friction force far away from the planet decays as $r^{-1}$ if
we integrate all the rings.  This explains why the force decays as
$t^{-1}$ in the subsonic case if we consider the slab geometry.
The rings that contribute to the dynamical friction
force reside at $r \sim ct$, as pointed out by \citet{Ost99}.

In this paper, we shall show in detail the derivation of the dynamical 
friction force exerted on the particle embedded in a gaseous slab using
the two-dimensional approximation.  Before going on 
to the details, we show that the dependences of the force on the
physical parameters of the gas can be derived in the two-dimensional
approximation.

In the realistic case of the gaseous slab with the thickness of 
$L_z (\sim \epsilon)$,
the interaction between the gas and the particle can be well
approximated by the two-dimensional analysis for the scales larger than
$L_z$ (far region).  Inside this region (near region), 
the interaction should be calculated in three-dimension.  
Integrating equation \eqref{estimate_3D} between the minimum cut-off
scale $r_{\rm min}$ and $\gamma L_z$ ($\gamma$ is a factor of the
order of unity), the contribution to the dynamical
friction force from the near region, $F_N$, is given by
\begin{equation}
 F_N \sim \dfrac{(GM)^2 \rho}{c^2} 
  \log \left( \dfrac{\gamma L_z}{r_{\rm min}} \right).
  \label{force_near}
\end{equation}
The contribution to the force from the far region, $F_F$, is given by
integrating equation \eqref{estimate_2D} from $\gamma L_z$ to infinity and, 
\begin{equation}
 F_F \sim \dfrac{(GM)^2 \Sigma}{c^2 \gamma L_z} 
  \sim \dfrac{(GM)^2 \rho}{\gamma c^2},
  \label{force_far}
\end{equation}
where we have used $\Sigma \sim \rho L_z$.  
The integral from the far region does not diverge at
$r=\infty$ since there is not enough mass to contribute to the force in
the slab geometry in the far region, in contrast to the case of the
homogeneous three-dimensional distribution of the gas.
It should be noted that the dependences of the force on physical
parameters in $F_F$ and $F_N$ are the same except for the logarithmic
factor, which only introduces very weak dependences on 
$L_z$ and $r_{\rm min}$.  
By choosing the appropriate value of $\gamma$, it is possible to have an
expression that would reproduce the results of three-dimensional
calculations from two-dimensional calculations.   
Therefore, we expect that the two-dimensional 
analyses can be used to understand the fundamental aspects of the
dynamical friction force.  
More detailed discussions on the
two-dimensional approximation, in relation to the application to the
disk-planet interaction, can be found in Section \ref{DF_2D_discuss}.

We now derive the steady state solution of equation \eqref{wave_lin}. 
We make use of the Fourier transform to solve the equations, 
but our goal is to derive the
expression of the force exerted on the particle, which is, of course,
the quantity in the real space.    
For any perturbed quantity $f(x,y)$, 
we define the Fourier transform by
\begin{equation}
 f(x,y) = \dfrac{1}{2\pi} \int dk_x dk_y \tilde{f}(k_x,k_y) 
  e^{i(k_x x + k_y y)},
  \label{FT_def}
\end{equation}
and the inverse transform is
\begin{equation}
 \tilde{f}(k_x,k_y) = \dfrac{1}{2\pi} \int dx dy f(x,y) 
  e^{-i(k_x x + k_y y)}.
  \label{invFT_def}
\end{equation}
The steady state of surface density perturbation can be derived 
from equation \eqref{wave_lin} in the Fourier space as
\begin{equation}
 \tilde{\alpha} = 
  - \dfrac{k^2 \tilde{\psi}}
  {c^2 k^2 - v_0^2 k_y^2 + i \nu v_0 k_y k^2},
  \label{denspert_sol}
\end{equation}
where $k^2 = k_x^2 + k_y^2$.  

If we know the profile of the perturbed surface density, 
the dynamical friction force exerted on the particle is calculated by
\begin{equation}
 F_y(x,y) = \int d^2x \delta \Sigma(x,y) \dfrac{\partial \psi}{\partial y}.
\end{equation}  
We note that the $x$-component of the force is zero because of the
symmetry.  From the solution in the  Fourier space, the dynamical
friction force is given by
\begin{equation}
 F_y(x,y) = 2 \int_0^{\infty} dk_y \int_{-\infty}^{\infty} dk_x
  k_y \tilde{\psi} \mathrm{Im} \left( \tilde{\alpha} \right),
  \label{force_Four}
\end{equation}
where the Fourier transform of the gravitational potential
$\tilde{\psi}$ is given by
\begin{equation}
 \tilde{\psi} = - \dfrac{GM}{k} e^{-k \epsilon} .
\end{equation}

We calculate the dynamical friction force using equations
\eqref{denspert_sol} and \eqref{force_Four}.  
By changing the integration variables
first from $(k_x,k_y)$ to $(k,\theta)$ via
\begin{equation}
 k_x = k \cos \theta
  \label{kx_trans}
\end{equation}
and 
\begin{equation}
 k_y = k \sin \theta,
  \label{ky_trans}
\end{equation}
and then from $k$ to $u=\nu v_0 k/c^2$, 
we can rewrite the integral as 
\begin{equation}
 F_y = \dfrac{\Sigma_0 (GM)^2}{c^2 \epsilon} 
  \dfrac{\Gamma}{2} 
  \int_0^{\infty} u I_{\theta} (M_0^2, u^2) 
  e^{ -\Gamma u }  du,
  \label{Force_int}
\end{equation}
where $M_0\equiv v_0/c$ is the Mach number of the background flow and 
$\Gamma\equiv 2c\epsilon/M_0 \nu$.  The
function $I_{\theta}(p,q)$ which appears in the integral is given by
\begin{equation}
 I_{\theta} (p,q) = \int_{0}^{2\pi} 
  \dfrac{\sin^2\theta}{(1-p\sin^2\theta)^2 + q \sin^2\theta} d\theta.
  \label{intI_expression}
\end{equation}
It is actually possible to perform the integration involved in
$I_{\theta}(p,q)$ analytically.  The explicit form of this is found in
Appendix \ref{app:intI}.

\subsection{Expressions in Some Limits}

It is now possible to perform the integral \eqref{Force_int} numerically
to obtain the dynamical friction force in a gaseous slab if we give two
dimensionless parameters $M_0$ and $\Gamma$.  
However, for a moment, 
we look at some limits to obtain analytic expressions in
order to investigate how the dynamical friction force depends on these
parameters. 

\subsubsection{Subsonic Limit}

We first look at the subsonic case where $M_0 \ll 1$.  In this case, we 
approximate $1-M_0^2 \sin\theta \sim 1$ so the integral
\eqref{intI_expression} becomes
\begin{equation}
 I_{\theta}(M_0^2,u^2) \sim \int_0^{2\pi} d\theta
  \dfrac{\sin^2\theta}{1+u^2 \sin\theta} .
\end{equation}
Therefore, in the limit of $u \to 0$, 
$I_{\theta}(M_0^2,u^2) \sim \pi$ while
in the limit of $u\to\infty$,
$I_{\theta}(M_0^2,u^2) \sim 2\pi/u^2$.  Connecting these two limits, we
approximate the integral by 
\begin{eqnarray}
 I_{\theta} (M_0^2,u^2) = 
  \left\{
   \begin{array}{cc}
       \pi & (u<\sqrt{2})  \\
     2\pi/u^2 & (u>\sqrt{2})
   \end{array}
  \right.
\end{eqnarray}
The dynamical friction force is obtained from equation
\eqref{Force_int}.  
It is possible to perform the integral analytically
and we have 
\begin{equation}
 F_y = \dfrac{\pi}{2} \dfrac{\Sigma_0 (GM)^2}{\epsilon c^2} \Gamma 
  \left[ \dfrac{1-(1+\sqrt{2}\Gamma)e^{-\sqrt{2}\Gamma}}{\Gamma^2}
  - 2 \mathrm{Ei}\left(-\sqrt{2}\Gamma \right) \right],
  \label{df_subsonic}
\end{equation} 
where $\mathrm{Ei}(x)$ is the exponential integral defined by
\begin{equation}
 \mathrm{Ei}(-x) = -\int_{x}^{\infty} dt \dfrac{e^{-t}}{t}.
\end{equation}
In the limit of subsonic perturbation, we expect that $\Gamma \gg 1$.
We then obtain
\begin{equation}
 F_y \sim \dfrac{\pi}{4} \dfrac{\Sigma_0 (GM)^2}{\epsilon c^2}
  \dfrac{M_0 \nu}{c \epsilon}.
  \label{df_subsonic_limit}
\end{equation}
The dynamical friction force is proportional to the speed of the
particle and also the viscosity.  
In the case of an inviscid, steady state with
a subsonic particle embedded in the gaseous slab, 
the dynamical friction force vanishes 
as indicated by the time-dependent analysis 
shown in Appendix \ref{app:timedep}.

\subsubsection{Supersonic Limit}

We now consider the supersonic limit where $M_0 \gg 1$.  
We first consider the behavior of the integral
$I_{\theta}(M_0^2,u^2)$ separately in the two limits, $u \ll M_0^2$ and
$u \gg M_0^2$.  

In the case of $u \ll M_0^2$, the dominant contribution to the integral
comes from $\theta \sim \theta_0$, where $\theta_0$ is given by
\begin{equation}
 \sin^2{\theta_0}=\dfrac{1}{M_0^2}.
\end{equation}
Approximating $\sin\theta=\sin(\theta_0+\delta \theta)$ to the first
order of $\delta\theta$,
\begin{equation}
 \sin (\theta_0+\delta \theta) \sim
  \sin\theta_0 + \delta \theta \cos\theta_0
\end{equation}
the integral can be approximated by
\begin{equation}
 I_{\theta}(M_0^2,u^2) \sim
  \dfrac{4}{M_0^2}
  \int_{-\infty}^{\infty} d \delta \theta 
  \dfrac{1}{4(M_0^2-1)\delta\theta^2 + (u/M_0)^2},
  \label{int_Lorentz}
\end{equation}
where the factor of $4$ comes from the fact that there are four
positions within $0< \theta <2\pi$ where $\sin^2\theta$ equals
$1/M_0^2$, and the range of integration is extended 
from $-\infty$ to $\infty$.  
The integrand of equation \eqref{int_Lorentz} is the Lorentzian function
with width $\sim u/M_0^2$.  The condition where this width must be very
small compared to the unity (or more exactly $2\pi$, in order to
justify the extension of the range of the integration) 
leads to the condition $u \ll M_0^2$.  
Using the formula
\begin{equation}
 \int_{-\infty}^{\infty} dx \dfrac{1}{1+x^2} = \pi,
\end{equation}
we obtain 
\begin{equation}
 I_{\theta}(M_0^2,u^2) \sim \dfrac{2\pi}{uM_0\sqrt{M_0^2-1}}.
\end{equation}

In the case of $u \gg M_0^2$, the width of the Lorentzian is too wide to
extend the integration range from $(0,2\pi)$ to $(-\infty,\infty)$.  In
this case, we expect that 
$u^2 \sin^2 \theta \gg (1-M_0^2\sin^2\theta)^2$
is satisfied in the most part of $\theta$.
We also approximate 
$1-M_0^2 \sin^2 \theta \sim - M_0^2 \sin^2\theta$ 
so we can rewrite the integral as
\begin{equation}
 I_{\theta}(M_0^2,u^2) \sim \int_0^{2\pi} 
  \dfrac{1}{M_0^2 \sin^2\theta + u^2}.
\end{equation} 
Using the formula
\begin{equation}
 \int_0^{2\pi} d\theta \dfrac{1}{a^2 + \sin^2 \theta}
  = \dfrac{2\pi}{a \sqrt{a^2+1}},
\end{equation}
we obtain
\begin{equation}
 I_{\theta}(M_0^2,u^2) \sim \dfrac{2\pi}{u^2},
\end{equation}
where we have used of the condition $u \gg M_0^2$.
In summary, in the supersonic limit, the integral 
\eqref{intI_expression} can be approximated by
\begin{eqnarray}
 I_{\theta}(M_0^2,u^2) =
  \left\{
  \begin{array}{cc}
    \dfrac{2\pi}{uM_0\sqrt{M_0^2-1}} & (u<M_0\sqrt{M_0^2-1}) \\
   \dfrac{2\pi}{u^2} & (u>M_0\sqrt{M_0^2-1})
  \end{array}
  \right.
\end{eqnarray}

We can now perform the integral \eqref{Force_int} to give the dynamical
friction force.  The result is
\begin{equation}
 F_y = \pi \dfrac{\Sigma_0 (GM)^2}{\epsilon c^2} \Gamma
  \left[ \dfrac{1-\exp\left(-\Gamma M_0 \sqrt{M_0^2-1}\right)}
   {\Gamma M_0 \sqrt{M_0^2-1}} - 
  \mathrm{Ei}\left( -\Gamma M_0 \sqrt{M_0^2-1} \right) \right].
  \label{df_supersonic}
\end{equation}
In the supersonic limit, we expect that $\Gamma M_0^2 \gg 1$.  In this
case the expression simplifies to
\begin{equation}
 F_y \sim \pi \dfrac{\Sigma_0 (GM)^2}{\epsilon c^2 M_0^2}.
  \label{df_supersonic_limit}
\end{equation}
As expected, we obtain that the force is proportional to $v_0^{-2}$.  

We note that the dynamical friction force is independent of the
viscosity in the supersonic limit.  
This indicates that the dynamical friction force in the supersonic limit 
is insensitive to the dissipation process.  
We conjecture that the dynamical friction force is
insensitive to the physical state of the slab (whether the slab is
isothermal or radiative) in the case of the supersonic motion. 
Similar situation happens in the discussion of the Lindblad torque
exerted on a planet embedded in a disk \citep{MVS87}.

\subsection{Dynamical Friction Force}

We now come to the point where we consider all the values of $M_0$.  We
integrate equation \eqref{Force_int} numerically to find the dependence
of the dynamical friction force on the physical parameters.  
As we have seen in the above analytic discussions, there are two
dimensionless parameters, 
$\Gamma$ and $M_0$, in the problem at hand.  
In this section, we use the ``Reynoldes number'', $Re$, defined by
\begin{equation}
 Re = \dfrac{c\epsilon}{\nu}
\end{equation}
instead of $\Gamma$, since the parameter $\Gamma$ contains both Mach
number and the viscous coefficient.  

\begin{figure}
 \plotone{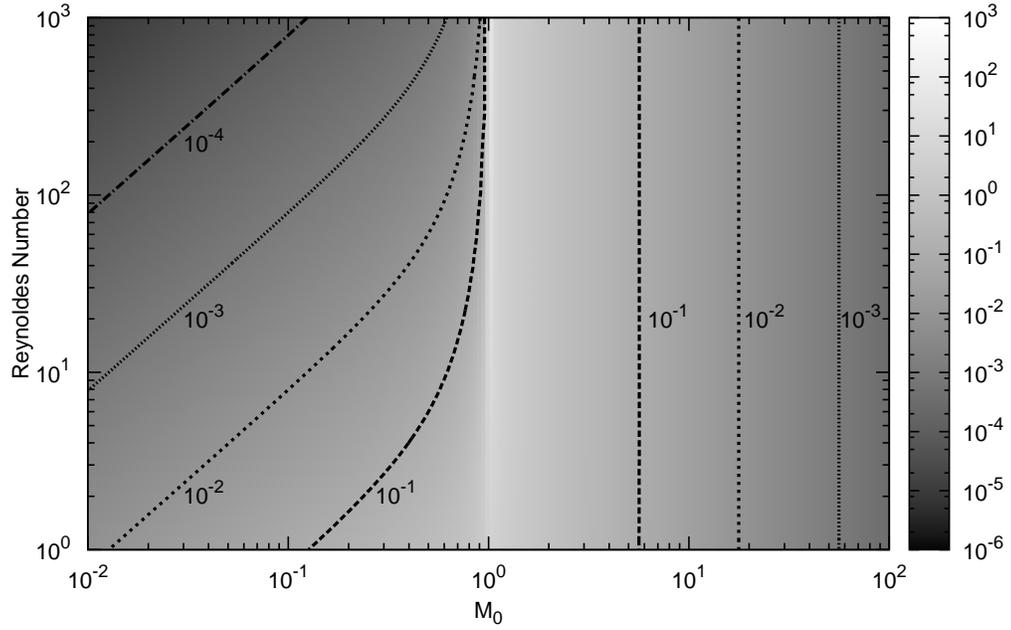}
 \caption{The dynamical friction force obtained by the numerical
 integration of equation \eqref{Force_int}.  The results are normalized
 by $\Sigma_0(GM)^2/c^2\epsilon$.  The horizontal axis shows the Mach
 number and the vertical axis shows the Reynoldes number (defined in the
 main text).   
 }
 \label{fig:df_re_beta}
\end{figure}

Figure \ref{fig:df_re_beta} shows the dependence of the dynamical
friction force normalized by $\Sigma_0 (GM)^2/c^2 \epsilon$ as a
function of Mach number and Reynoldes number.  In the subsonic regime,
we see the expected behavior from equation \eqref{df_subsonic_limit}
where dynamical friction force decreases as we decrease the Mach number
and viscosity.  For the supersonic case, we also see the behavior
expected from equation \eqref{df_supersonic_limit} where the force
decreases as we increase the velocity but the force only weakly depends
on the values of viscosity.  

\begin{figure}
 \plotone{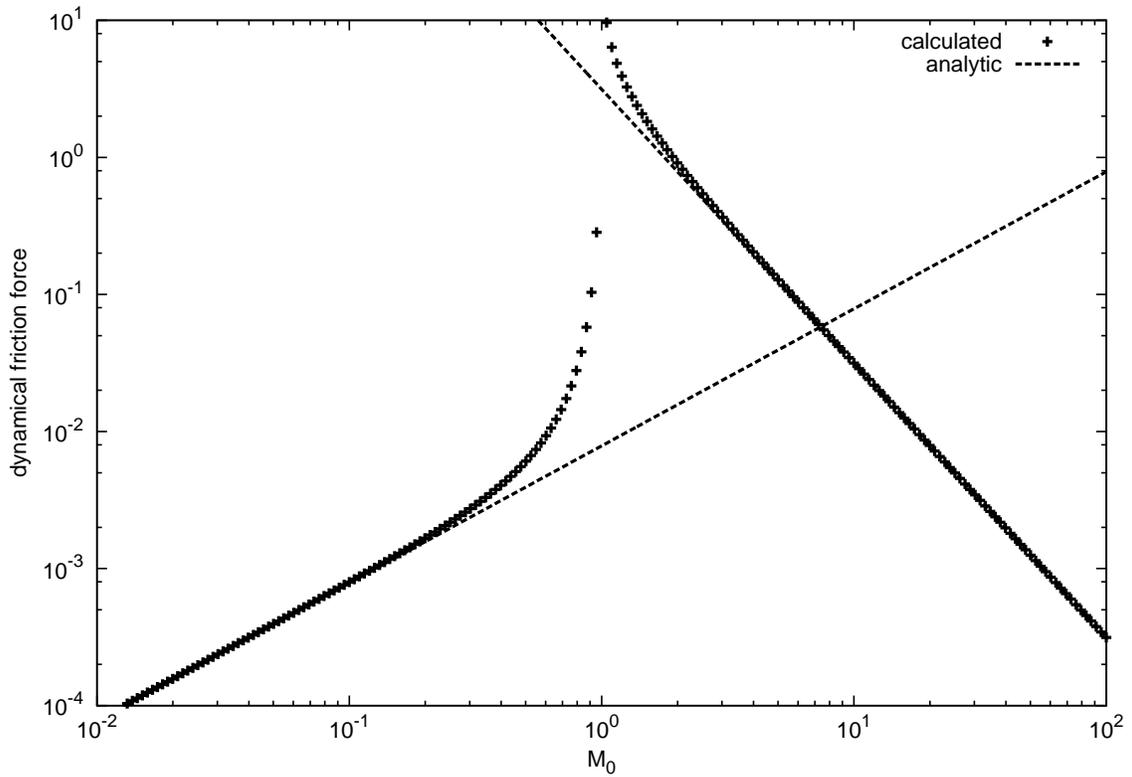}
 \caption{The dynamical friction force for $Re=100$.  The results of
 numerical integration of equation \eqref{Force_int} are shown by
 plus symbols, and the analytic formulae in the limit of supersonic and
 subsonic cases, which are given by equations \eqref{df_subsonic_limit}
 and \eqref{df_supersonic_limit} respectively, are shown by dashed
 lines.   
 }
 \label{fig:df_comp_re100}
\end{figure}

Figure \ref{fig:df_comp_re100} shows the dependence of the dynamical
friction force on the Mach number in the case of $Re=100$, and we also
show the formulae in the subsonic and supersonic limit, equations
\eqref{df_subsonic_limit} and \eqref{df_supersonic_limit}.  These
limiting formulae well describe the behaviors of the dynamical friction
force.  Equation \eqref{df_supersonic_limit} is actually 
a good approximation of the dynamical friction force even in the case of
$M_0\sim 2$.

We briefly comment on the divergence at $M_0=1$.  In Figures
\ref{fig:df_re_beta} and \ref{fig:df_comp_re100}, we see that the
dynamical friction force diverges at $M_0=1$.  This divergence comes
from the matching between the background velocity and the sound speed,
and was already seen in the previous linear analyses by \citet{Ost99}.
It is not possible to avoid this divergence by the effect of viscosity. 
We consider that nonlinear effects are important here.

\subsection{Validity of 2D Approximation}
\label{DF_2D_discuss}

We have derived the dynamical friction force exerted on a particle
embedded in a homogeneous gas slab using the two-dimensional
approximation.  We now discuss the validity of this
approximation in the subsonic and supersonic cases.  

As we have discussed before, the two-dimensional treatment is based on
the averaging of the equations in the vertical direction, and the
phenomena that occur on the scales larger than the vertical averaging
scale can be well approximated by the two-dimensional calculations.  
In our model, the vertical averaging scale is given by $\epsilon$, which
is the softening scale of the gravitational potential in equation
\eqref{pot_2D}.  

In the subsonic case, the dominant contribution to the integral
\eqref{Force_int} comes from $\Gamma u \sim \mathcal{O}(1)$, which is 
$\epsilon k \sim \mathcal{O}(1)$.  This indicates that in the subsonic
 case, the perturbation with the scale comparable to the gravitational 
 softening length becomes important in determining the dynamical
 friction force.  
Therefore, it is indicated that the full three-dimensional treatment may
be necessary to have a more quantitative results.  

In the case of the supersonic motion, let us consider the case where 
$\Gamma M_0^2 = 2 v_0 \epsilon/\nu \gg 1$ for simplicity.  
In this case, the integral \eqref{Force_int} 
for the dynamical friction force is approximated by
\begin{equation}
 F_y \propto \int_{0}^{\infty} du \dfrac{e^{-\Gamma u}}{M_0^2}
  \propto \int_{0}^{\infty} dk e^{-\epsilon k},
\end{equation}
and therefore, all the scales satisfying  
$\Gamma u \sim \epsilon k  \lesssim 1$
contribute to the force.  In other words, all the scales 
from the large scale (small $k$), 
where we expect that the two-dimensional
approximation is valid, down to the cutoff scale contribute to the
force.  
Since the dependence of the cut-off scale $\epsilon$ is also present in
the case of the supersonic motion, rigorous three-dimensional treatment
is necessary to give more quantitative expressions of the dynamical
friction force.  

However, we expect that it is possible to capture the
dependences on physical parameters even in the two-dimensional 
approximation.  
In the beginning of Section \ref{dynamicalfric_lin}, we
have estimated the order of the magnitude of the dynamical
friction force in equations \eqref{estimate_3D} and \eqref{estimate_2D},
and the contribution from the near region \eqref{force_near} and that
from the far region \eqref{force_far} differ by only a logarithmic
factor $\log (L_z/r_{\rm min})$.   
The dynamical friction force derived in this paper corresponds to the
contribution from the far region.
The lack of the logarithmic factor is due to the fact that the
potential is smoothed in the region $r \lesssim \epsilon$ (see also the
discussion after equation \eqref{pot_2D}).  
We note the similarity of the expressions derived by the simple
order-of-magnitude estimate \eqref{force_far} 
and more rigorous calculations \eqref{df_supersonic_limit}.

In order to estimate the contribution from the near region, it is
necessary to determine the appropriate value of $r_{\rm min}$.
However, it is not straightforward and 
there are several publications on this using the non-linear 
calculations \citep[e.g.,][]{SSB99,KK09}.  
Let us consider, for a moment, the case of the planet embedded in a
protoplanetary disk.  
Naive estimate of $r_{\rm min}$ is the radius of the planet, and in the
case of an Earth-mass planet, it is of the order of $10^4\mathrm{km}$.  
If we use the typical disk scale height $H \sim 0.05\mathrm{AU}$, the
value of $\log(H/r_{\rm min})$ amounts to $\sim 7$. 
Recent non-linear calculations \citep{KK09} suggest that the deviation
from the linear value may be smaller due to the shock formation.  
If we use $r_{\rm min} \sim GM/v^2$, 
which is the typical scale of the distance between the particle and the
shock in supersonic motion \citep[e.g.,][]{KK09}, 
instead of the planet radius for $r_{\rm min}$, 
the logarithmic factor is $\sim 3-6$ for an Earth-mass body. 
We shall give further discussions on the non-linear
effects on dynamical friction at the end of Section \ref{sec:diskplanet}.  
\footnote{ 
The use of this value for $r_{\rm min}$ instead of the radius of the
body is also motivated by the results of the collisionless system, where
$90^{\circ}$ deflection angle appears as $r_{\rm min}$. }
In any case, the force derived in this section may be different by some 
factor due to the two-dimensional approximations,   
but such logarithmic factor can be approximately taken into account by 
choosing the appropriate value for $\epsilon$ (or $\gamma$ in equation
\eqref{force_far}).

Despite the uncertainty of the two-dimensional approximation,
especially for the value of $\epsilon$ that should be taken,
we still apply this results
to the problem of the disk-planet interaction.  It is because many of
the calculations to date are done in two-dimensional approximations and
they involve the potential of the form given in equation \eqref{pot_2D}.  
One goal of this paper is to investigate how well the simple model using
the dynamical friction force may be able to describe the numerical
results of the disk-planet interaction.  More rigorous treatment of the
dynamical friction force including the three-dimensional effects and the
vertical stratification will be discussed in future
publications.  It is also necessary to have three-dimensional
non-linear calculations of the disk-planet interaction to compare the
model.  We note that the discussion on the form of the potential applies
not only to the linear perturbation analyses, but also to the non-linear
calculations.

\section{Gravitational Interaction between a Disk and an Eccentric Planet}
\label{sec:diskplanet}

In this section, we apply the results of the dynamical friction obtained
in the previous section to the problem of disk-planet interaction.  
In this paper, we particularly focus on the planet in a highly eccentric
orbit. 

Before going on to the main topic, we briefly summarize the notation and
the terminology.  In this paper, we
focus on the evolution of the semimajor axis $a$ and the eccentricity
$e$ of the planet.  We denote the timescale of the evolution of the
semimajor axis by $t_a$, which is defined by 
\begin{equation}
 t_a = -\dfrac{a}{\overline{da/dt}},
  \label{ta_def}
\end{equation}
where $\overline{da/dt}$ is the time derivative of the osculating
element averaged over one orbital period.  
Similarly, we denote the evolution timescale of the eccentricity by
$t_e$, which is defined by 
\begin{equation}
 t_e = \dfrac{e}{\overline{|de/dt|}}.
  \label{te_def}
\end{equation}
In \citet{PL00}, they use ``migration timescale'' $t_m$, which is
defined by
\begin{equation}
 t_m = - \dfrac{J}{\overline{dJ/dt}},
  \label{tm_def}
\end{equation}
where $J$ is the specific angular momentum of the planet.  
It is noted that $t_a$ and $t_m$ are not the same, 
\footnote{It is noted that $t_a$ and $t_m$ are different by a factor of
two even in the case of the circular orbit.}
since the angular
momentum $J$ is given by 
\begin{equation}
 J = \sqrt{a(1-e^2)}.
\end{equation}
However, it is possible to express $t_m$ in terms of $t_a$ and $t_e$.
In this paper, we mainly use $t_a$ and $t_e$, which we refer to as
``semimajor axis evolution timescale'' and ``eccentricity damping
timescale'', respectively.  
We use $t_m$ from time to time when necessary, and referred to it as
``migration timescale'', but it should not be confused with the
semimajor axis evolution timescale.  

The gravitational interaction between a planet and a protoplanetary disk
has been investigated by many authors.  
The standard formulation for the linear perturbation analysis of the 
disk-planet interaction is done as follows \citep{GT79,GT80,Art93,PL00}.  
The planetary orbit is decomposed into the power series of eccentricity,
and then the planetary potential is decomposed into the Fourier series
in the azimuthal direction.  For each Fourier component, the
perturbation excited by the planetary potential at resonances (Lindblad
resonances and corotation resonances) are calculated.  Then, the
contribution from all the resonances are summed up to obtain the force
exerted on the planet, which is readily applied to obtain the orbital
evolution of the planet.  

\citet{PL00} calculated the torque and energy exchange between a
planet and a disk using this formalism and obtained the migration
timescale and the eccentricity damping timescale.  
They have found that the eccentricity always damps.
They have obtained the 
migration rate and the eccentricity damping timescale as
\begin{equation}
 t_m = 3.5 \times 10^5 f_s^{1.75} 
  \left[  
   \dfrac{1+(er_0/1.3H)^5}{1-(er_0/1.1H)^4}
  \right]
     \left( \dfrac{H/r_0}{0.07} \right)^2
   \left( \dfrac{2M_{\rm J}}{M_{\rm GD}} \right)
   \left( \dfrac{M_{\oplus}}{M_{\rm p}} \right)
   \left( \dfrac{r_0}{1\mathrm{AU}} \right)
   \mathrm{yr}
   \label{PL_mig}
\end{equation}
and
\begin{equation}
 t_e = 2.5 \times 10^3 f_s^{2.5} 
  \left[  
   1 + \dfrac{1}{4} 
   \left(\dfrac{e}{H/r_0}\right)^3
  \right]
     \left( \dfrac{H/r_0}{0.07} \right)^4
   \left( \dfrac{2M_{\rm J}}{M_{\rm GD}} \right)
   \left( \dfrac{M_{\oplus}}{M_{\rm p}} \right)
   \left( \dfrac{r_0}{1\mathrm{AU}} \right)
   \mathrm{yr}.
   \label{PL_e}
\end{equation}
Here, $H$ is the scale height of the disk, $r_0$ is the semimajor axis
of the planet, $M_{\rm GD}$ is the disk mass contained in $5\mathrm{AU}$,
$M_{\rm p}$ is the planet mass.  The parameter $f_s$ is related to the
softening length $\epsilon$ of the planet's gravitational potential by
$f_s=(2.5\epsilon/H)$.  
They performed the analysis for the disk with the 
constant aspect ratio, $H/r$, and the surface density variation with
$r^{-3/2}$.  
From these equations, the eccentricity damps exponentially when $e \ll 1$,
while the damping timescale is proportional to $e^3$ for $e \gtrsim H/r$

The three-dimensional linear perturbation analysis for a planet in an
eccentric orbit is done by \citet{TW04}.  They have found, for a small
eccentricity $e \ll 1$, the eccentricity damps exponentially as
\begin{equation}
 t_e = 1.282 \dfrac{M_{\ast}}{M_{\rm p}}
  \dfrac{M_{\ast}}{\Sigma r_0^2} 
  \left( \dfrac{H}{r_0} \right)^4
  \Omega_{\rm p}^{-1},
  \label{TW_edamp}
\end{equation}
where $M_{\ast}$ is the mass of the central star and $\Omega_{\rm p}$ is
the angular frequency of the planet.  
If we adopt 
$\Sigma = 6 \times 10^{2} (r/1\mathrm{AU})^{-3/2} \mathrm{g/cm}^2$, 
which corresponds to the model where the mass contained within
$5\mathrm{AU}$ is equal to $2M_J$, $H/r_0=0.07$, 
and $M_{\rm p}=M_{\oplus}$, equation \eqref{TW_edamp} gives
$t_e \sim 2.4\times 10^{4}\mathrm{yr}$ at $r_0=1\mathrm{AU}$.
This value seems one order of magnitude larger than equation
\eqref{PL_e} with $f_s=1$.  
However, given the uncertainty of the softening length, 
it may be possible to obtain the consistent results if one adopts
$\epsilon \sim H$.  It is necessary to do a detailed comparison between 
2D and 3D calculations to derive the reasonable values of the softening
length.

The disk-planet interaction when the planet was in an eccentric orbit was 
investigated by using non-linear numerical simulations by 
\citet{CN06,CDKN07} and more recently by \citet{BK10} using a fully
radiative code.  They observed that when the eccentricity was smaller
than $0.1$, the eccentricity would damp exponentially, and the formula
by \citet{TW04} was in good agreement with the numerical results.  For
higher eccentricity such as $e \sim 0.3-0.4$, they observed that 
the eccentricity damping would slow down, 
and the timescale was in excellent agreement
with the formula presented by \citet{PL00}.  
\citet{BK10} found that for a planet with high eccentricity
($e\sim0.4$), fully radiative calculations and locally isothermal
calculations would give the same results.  
For the migration timescale, \citet{CN06} reported 
that the formula given by 
\citet{PL00} consistently predicted the migration timescale three times 
shorter for the planets with small eccentricity, while for those with
large eccentricity, the formula consistently predicted the timescale
$1.5$ times faster.  

So far, the agreement between the numerical calculations and linear
analysis is good.  However, since this formulation involves the
expansion of the planetary orbit 
in a power series of eccentricity $e$, these formulae are applicable to
the case where $e \ll 1$. 
It is noted that the formula by \citet{PL00} is 
applicable to the case where $e \gtrsim H/r_0$ but $e \ll 1$ is still
necessary.    

In this paper, we present an alternative model for the disk-planet
interaction with a planet in an eccentric orbit, which is especially
useful for a planet with high eccentricity.  We model the interaction
using the dynamical friction formula and our calculation proceeds as
follows.   
We first calculate the relative velocity
between the gas and the planet.  Then, we make use of the dynamical
friction formula to obtain the force exerted on the planet at each
location of the orbit.  From these forces, we obtain the evolution of
the orbital semimajor axis and eccentricity using Gauss's equations, 
and when averaged over the orbital period of the planet, 
we finally obtain the timescale for the evolution of the orbital
parameters.  In the following, we describe each step one by one.
We shall show that the timescales of the evolution of the orbital
parameters obtained in this way are in good agreement with the previous
work, although the instantaneous force exerted on the planet on each
location of the orbit may be rather over-simplified.  
We show how such timescales vary as we consider various disk models.  
We note that the usefulness of the dynamical friction formula in the
problem of disk-planet interaction was hinted in \citet{P02}.  However,
the results were shown for some limited number of disk models.

The big assumption in this model is that we neglect the effects of
local shear at the location of the planet.  
Later in this section, we discuss the applicability of this formulation
in conjunction with this assumption.

\subsection{Setup of the Problem}

We consider a protoplanetary disk with a central star with mass
$M_{\ast}$ and surface density profile with $r^{-p}$,
\begin{equation}
 \Sigma = \Sigma_0 \left( \dfrac{r}{r_0} \right)^{-p},
\end{equation}
where $\Sigma_0$ is the surface density at $r=r_0$.
We assume for simplicity that the disk is locally isothermal with the
temperature profile $T \propto r^{-q}$.  Then, the sound speed $c$ of
the disk gas varies as $c \propto r^{-q/2}$.  
We assume that the disk is rotating at the Kepler angular frequency and
neglect the small difference arising from the pressure gradient.  We
define the scale height of the disk $H$ by $H=c/\Omega_{\rm K}$,
where $c$ is the sound speed and $\Omega_{\rm K}$ is the Kepler angular
frequency.  The scale height $H$ is therefore varies as 
$H \propto r^{(3-q)/2}$.  The disk aspect ratio, $H/r$, is given by 
\begin{equation}
 \dfrac{H}{r} = h_0 \left( \dfrac{r}{r_0} \right)^{(1-q)/2},
\end{equation}  
where $h_0$ is the disk aspect ratio at $r=r_0$.
We consider a planet with mass $M_{\rm p}$ with semimajor axis $a$ and
eccentricity $e$.  

In the later sections, we shall often refer to the ``fiducial model'', 
which we define as $p=3/2$, $q=1$, $M_{\ast}=M_{\odot}$, 
$M_{\rm p}=M_{\oplus}$, $h_0=0.05$, and
$\Sigma_0$ is chosen in such a way that the mass contained within
$5\mathrm{AU}$ is $2M_{\rm J}$.  In this model, the disk aspect ratio is
constant throughout the disk.  We call this model ``fiducial'' because
it is the model used in the numerical simulations by \citet{CN06}, 
which we shall compare our analytic results with their numerical
calculations.  
Note that this is not exactly the same as Minimum Mass Solar Nebula.

The gravitational potential of the planet is given by 
equation \eqref{pot_2D}, where $(x,y)$ is now the coordinate
centered on the planet.  
The softening length $\epsilon$ is given by 
$\epsilon=\epsilon_0 H(r)$ where $\epsilon_0$ is the dimensionless
parameter.  \citet{CN06} used $\epsilon=0.5$.  We note that in our
model, the softening parameter varies as the location of the disk if $H$
varies as $r$.

\subsection{Relative Velocity between the Gas and the Planet}

We first calculate the relative velocity between the planet and the
disk.   
It is readily calculated if we give the planet's semimajor axis and
eccentricity, since we assume the disk is in Keplerian rotation.  

\begin{figure}
 \plotone{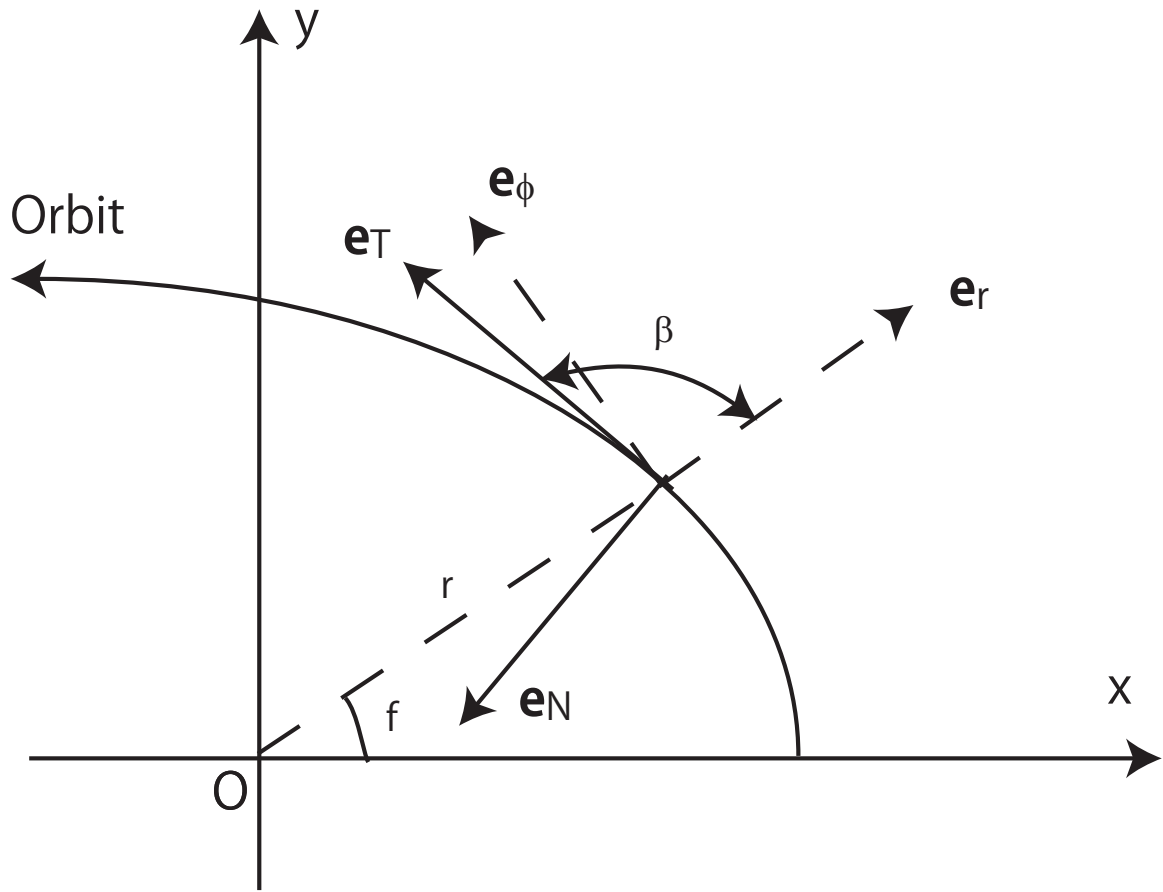}
 \caption{Definition of unit vectors $\mathbf{e}_T$ and $\mathbf{e}_N$
 and their relations with the unit vectors in the cylindrical coordinate
 $\mathbf{e}_r$ and $\mathbf{e}_{\phi}$.  The angle between
 $\mathbf{e}_r$ and $\mathbf{e}_T$ is $\beta$.
 }
 \label{fig:coord}
\end{figure}

We assume that the planet mass $M_{\rm p}$ is negligible compared to the
mass of the central star $M_{\ast}$ and use 
the cylindrical coordinate system $(r,\phi,z)$ with the origin at
the central star.  
We denote the mean motion of the planet by $n=(GM_{\ast}/a^3)^{1/2}$, 
and the true anomaly by $f$.  The velocity of the planet is then 
\begin{equation}
 \mathbf{v}_{\rm p} = \dfrac{aen}{\eta} \sin f \mathbf{e}_r
  + \dfrac{an}{\eta} \left( e\cos f +1 \right)\mathbf{e}_{\phi},
\end{equation}
where $\mathbf{e}_r$ and $\mathbf{e}_{\phi}$ are the unit vectors in
$r$- and $\phi$-directions, respectively.  The velocity of the gas is
\begin{equation}
 \mathbf{v}_{g} = \sqrt{\dfrac{GM_{\ast}}{r}} \mathbf{e}_{\phi}
  = \sqrt{\dfrac{n^2 a^3}{r}} \mathbf{e}_{\phi}.
\end{equation}
The relative velocity between the gas and the planet is calculated by 
$\Delta \mathbf{v}= \mathbf{v}_g-\mathbf{v}_{\rm p}$.
Now, we define the unit vectors $(\mathbf{e}_{T},\mathbf{e}_N)$, where
$\mathbf{e}_T$ is directed towards the velocity of the planet and
$\mathbf{e}_N$ is perpendicular to $\mathbf{e}_T$ and in the orbital
plane (see Figure \ref{fig:coord}).  Unit vectors 
$(\mathbf{e}_r,\mathbf{e}_{\phi})$ and
$(\mathbf{e}_T,\mathbf{e}_N)$ are related by
\begin{equation}
 \mathbf{e}_r = \mathbf{e}_T \cos \beta - \mathbf{e}_N \sin\beta
\end{equation}
\begin{equation}
 \mathbf{e}_{\phi} = \mathbf{e}_T \sin \beta + \mathbf{e}_N \cos\beta, 
\end{equation}
where 
\begin{equation}
 \cos \beta = \dfrac{e \sin f}{\sqrt{1+2e\cos f + e^2}}
\end{equation}
and
\begin{equation}
 \sin \beta = \dfrac{1 + e \cos f}{\sqrt{1+2e\cos f + e^2}}.
\end{equation}

\begin{figure}
 \plotone{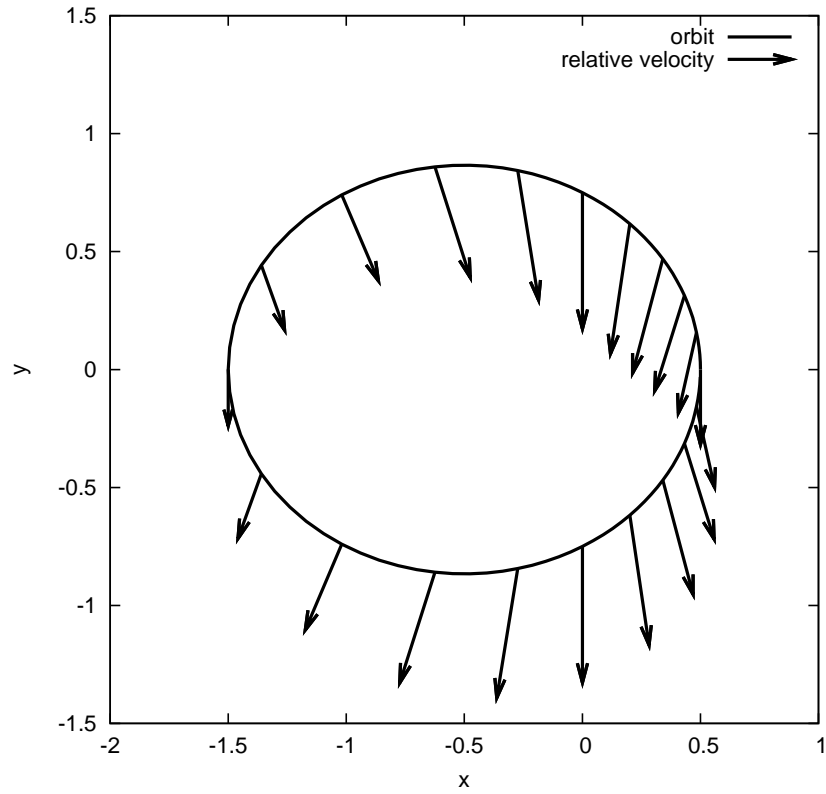}
 \caption{The evolution of relative velocity vector along the planetary
 orbit with $e=0.5$.  Solid line shows the orbit of the planet, and the
 arrows show the direction of relative velocity vector between the gas
 and the planet.  We have assumed that the central star is at the
 origin, and the planet and the disk are both
 rotating in the counterclockwise direction.
 }
 \label{fig:relv_orb_e05}
\end{figure}

\begin{figure}
 \plotone{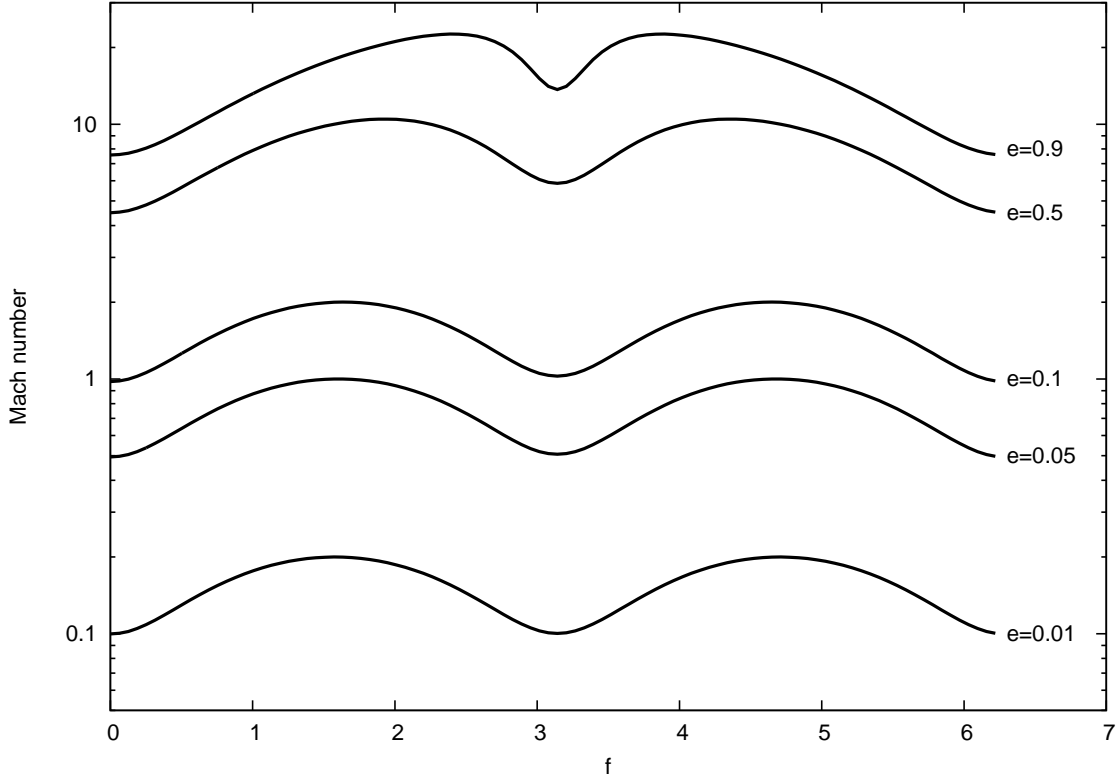}
 \caption{The amplitude of relative velocity between the gas and the
 planet over one period.  Horizontal axis shows the true anomaly $f$,
 and the vertical axis shows the Mach number of the relative speed.  The
 perihelion is at $f=0$ and the aphelion is at $f=\pi$.  
 The planet with eccentricity $e=0.01,0.05,0.1,0.5,0.9$ is shown.  
 We use the fiducial model (see main text) for the gas disk.  
 }
 \label{fig:relv_vare}
\end{figure}

In Figure \ref{fig:relv_orb_e05}, we show the evolution of the relative
velocity vector $\Delta \mathbf{v}$ over one orbital period with
$e=0.5$.  The planet feels headwind at the perihelion, and
tailwind at the aphelion.  Figure \ref{fig:relv_vare} shows the
amplitude of the relative velocity over one orbital period.  
It is noted that if the planet's eccentricity exceeds the disk aspect
ratio, the relative velocity is always supersonic, regardless of the
position in the orbit.

\subsection{Orbital Evolution of the Planet: Calculation Methods}

We make use of the Gauss's equations to calculate the orbital evolution
of the planet.  For the evolution of semimajor 
axis and eccentricity, Gauss's equations read
\begin{equation}
 \dfrac{da}{dt} = \dfrac{2 v}{a n^2} T
  \label{Gauss_a}
\end{equation}  
and
\begin{equation}
 \dfrac{de}{dt} = \dfrac{1}{v} 
  \left[  
   2(e + \cos f)T - \dfrac{r}{a} \sin f N
  \right],
  \label{Gauss_e}
\end{equation}
where $v$ is the speed of the planet and  
$r$ is the distance between the planet and the central star.  
We denote the perturbing force per unit mass 
by $(T,N)$, where $T$ is the component of the force (per unit mass) 
in the tangential direction of the orbit (positive direction is the
direction of the velocity), and $N$ is the component in the plane of the
orbit and perpendicular to $T$ (positive direction is directed inside
the orbital ellipse).  The evolution equation for the true anomaly is
given by
\begin{equation}
 \dfrac{df}{dt} = \dfrac{a^2 n \eta}{r^2} 
  + \dfrac{\eta}{nae} 
  \left[
   R \cos f - S \left( 1+\dfrac{a(1-e^2)}{r} \sin f \right)
  \right],
  \label{Gauss_f}
\end{equation}
where $\eta=(1-e^2)^{1/2}$, $R$ is the
component of the force in the radial direction, and $S$ is the component
perpendicular to $R$.  Components $(T,W)$ and $(R,S)$ are related by
\begin{equation}
 R = T \cos \beta - N \sin \beta
\end{equation}
and
\begin{equation}
 S = T \sin \beta + N \cos \beta.
\end{equation}

For the expressions of the perturbing force $(T,W)$, we use the dynamical
friction formula.  In this paper, we especially focus on the planet with
high eccentricity, and therefore assume that 
the relative speed between the gas and the planet is always supersonic.  
Therefore, we use the limiting form of highly supersonic case, equation
\eqref{df_supersonic_limit}.  The amplitude of the force is 
\begin{equation}
 F = \dfrac{\pi \Sigma(r) G^2M_{\rm p}}{\epsilon \Delta v^2},
  \label{df_model}
\end{equation}  
where $\Delta v$ is the relative speed between the gas and the
planet.  The direction of the force is opposite to the relative velocity
vector.  

In order to find the long-term evolution of orbital parameters, 
we average equations \eqref{Gauss_a}, \eqref{Gauss_e}, and
\eqref{Gauss_f} over one orbital period assuming the planet is in a
fixed orbit.  Since the variation of $a$ and $e$ is very small over one
orbital period, the assumption of a fixed orbit is justified.  The
averaging is done in such a way that
\begin{equation}
 \overline{\dfrac{da}{dt}} = \dfrac{1}{T_{\rm orb}} 
  \int_0^{T_{\rm orb}} dt \dfrac{da}{dt}
  = \dfrac{1}{T_{\rm orb}} \int_0^{2\pi} df \dfrac{da/dt}{df/dt}, 
\end{equation}
where $T_{\rm orb}$ is the time taken over one period.
The averaging is done in the same way for eccentricity.

\subsection{Orbital Evolution of the Planet: Fiducial Model}

In this section, we compare our results with previous numerical
calculations for the fiducial model.

\begin{figure}
 \plottwo{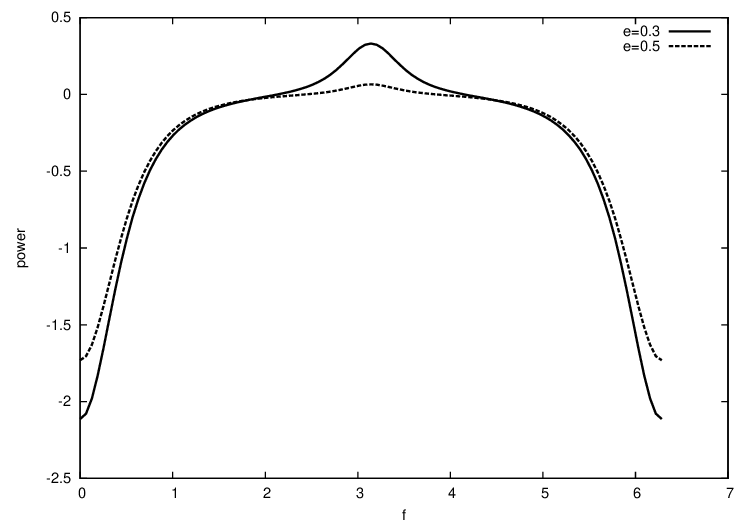}{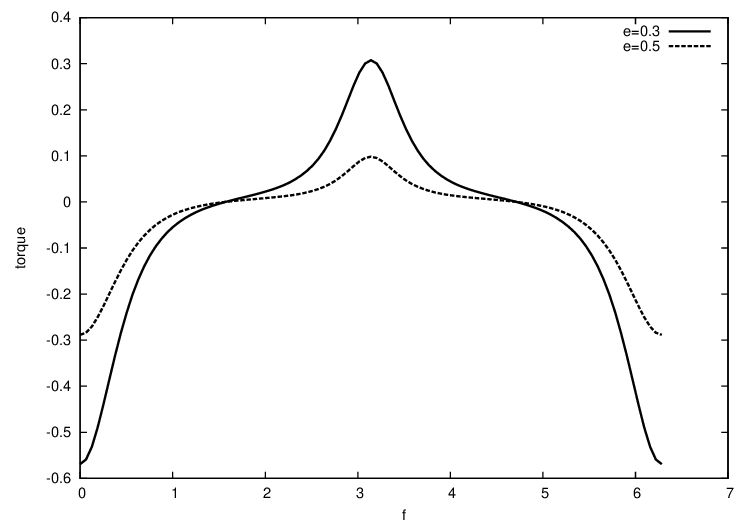}
 \caption{The power $\mathcal{P}$ (left panel) and torque $\mathcal{T}$
 (right panel) exerted on the planet with eccentricity $e=0.3$ (solid
 line) and $e=0.5$ (dashed line).  The fiducial model is used.  For the
 power, we plot $(\mathcal{P}/n E_{\rm p})/(M_{\rm p}/M_{\ast})$ and for 
 the torque, 
 we plot $(\mathcal{T}/n L_{\rm p})/(M_{\rm p}/M_{\ast})$, where 
 $E_{\rm p}$ and $L_{\rm p}$ are the planet's energy and angular
 momentum, respectively, and $n$ is the mean motion of the planet.  The
 horizontal axis shows the true anomaly $f$, and $f=0$ corresponds to
 the perihelion.
 }
 \label{fig:powertorque}
\end{figure}

We first show the torque and power exerted on the planet by the disk
over one orbit.  The power $\mathcal{P}$ 
exerted on the planet is given by 
\begin{equation}
 \mathcal{P} = \mathbf{v}_{\rm p} \cdot \mathbf{F}_{\rm disk} = vT
\end{equation}
and the torque $\mathcal{T}$ is 
\begin{equation}
 \mathcal{T} = \mathbf{e}_z \cdot \left(\mathbf{r}_{\rm p} 
  \times \mathbf{F}_{\rm disk}\right) = rS, 
\end{equation}
where $\mathbf{r}_{\rm p}$, $\mathbf{v}_{\rm p}$, 
and $\mathbf{F}_{\rm disk}$ are the position vector of the planet from
the disk, the velocity of the planet, and the force exerted on the
planet, respectively. 
In Figure \ref{fig:powertorque}, we show $\mathcal{P}$ and $\mathcal{T}$
normalized by the planet's energy (the sign inverted) 
\begin{equation}
 E_{\rm p}= \dfrac{1}{2}\dfrac{GM_{\ast}}{a} 
\end{equation} 
and the angular momentum 
\begin{equation}
 L_{\rm p}= \sqrt{GM_{\ast}a \left( 1-e^2 \right)}, 
\end{equation}
respectively.

In the vicinity of the perihelion, the torque and the power are negative
because of the headwind towards the planet, 
and they are positive in the vicinity of the aphelion owing to the
tailwind.  
Similar behavior is observed in the numerical simulations of
\citet{CDKN07}.  
Especially, the behavior of the torque exerted on the planet looks very
similar except for the small difference between the location of the 
perihelion or aphelion and the location of the maximum or minimum of the
torque (see Figure 8 of \citet{CDKN07}).   
In our model, this small difference in phase does not appear since we
assume that the force is proportional to the relative velocity at the
location of the planet.  

\begin{figure}
 \plottwo{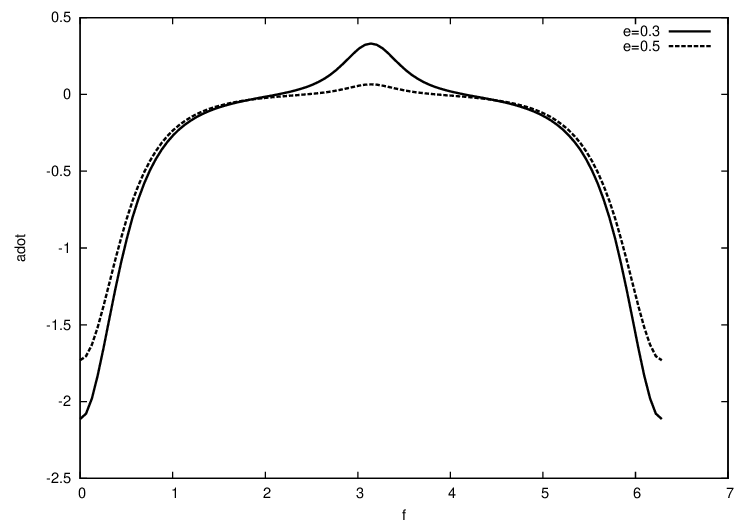}{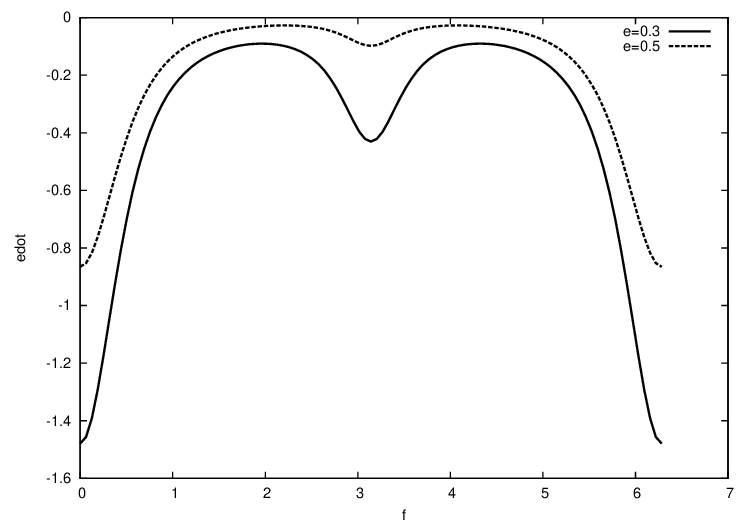}
 \caption{The evolution of semimajor axis $da/dt$ (left panel) 
 and eccentricity $de/dt$ (right panel) 
 of the planet with eccentricity $e=0.3$ (solid line) 
 and $e=0.5$ (dashed line).  The fiducial model is used.  
 For the evolution of semimajor axis, we plot
 $(da/dt)(1/na)/(M_{\rm p}/M_{\ast})$ and for
 the evolution of eccentricity, 
 we plot $(de/dt)(1/n) / (M_{\rm p}/M_{\ast})$.  The
 horizontal axis shows the true anomaly $f$, and $f=0$ corresponds to
 the perihelion.
 }
 \label{fig:adotedot}
\end{figure}

The comparison of our results with the Figure 10 of \citet{CDKN07} shows 
that the behavior of the power is very different.  
This difference can not be attributed to the different sets of
parameters they have used from our fiducial model.  
This may be because our model for the force exerted on the planet is
very simplified.  This difference in power leads to the different
behavior in $da/dt$ and $de/dt$ over one orbital period, which are shown
in Figure \ref{fig:adotedot}.  
We note here that the evolution of eccentricity is
qualitatively different from the Figure 9 of \citet{CDKN07}.
Fortunately, however, it is possible to obtain quantitatively the
same results with previous numerical calculations when we take the
average over one orbital period, as shown below.

Figure \ref{fig:tmtecomp} shows the results of migration rate $t_m$ and
the eccentricity damping rate $t_e$, which are obtained by the averaging
over one orbital period.  
Also plotted in this figure are the formula
obtained by \citet{PL00} given in equations \eqref{PL_mig} and
\eqref{PL_e}.  For the migration rate, we also show $t_m$ given by
\citet{PL00} times $3/2$ for $e<0.5$, 
which describes the results of numerical simulations better as
noted by \citet{CN06}.  For the damping of eccentricity, the formula by
\citet{PL00} fits the results of numerical simulations well.  
We see that our results on $t_e$ agree very well with the formula by
\citet{PL00} for $e \lesssim 0.7$.  For the migration rate, our
formulation consistently results in a factor of two slower timescale
compared to the numerical simulations.  Our results deviate from those
of \citet{PL00} for large values of eccentricity.  We expect that our
treatment is actually better for highly eccentric cases (discussed in
later sections), 
and than this difference is attributed to the breakdown of
the expansion of $e$ used by \citet{PL00}.

\subsection{Orbital Evolution of the Planet: Varying Disk Model}

We have seen that our formulation results in the consistent timescale of
the evolution of orbital parameters for the fiducial model.  We now look
at how the timescales behave as we vary the disk model.

In our framework, the force exerted on the planet can be written in the
form, 
\begin{equation}
 T = K r^{-\alpha} \dfrac{\Delta v_T}{\Delta v^3}
\end{equation}
and
\begin{equation}
 N = K r^{-\alpha} \dfrac{\Delta v_N}{\Delta v^3},
\end{equation}
where $K$ and $\alpha$ are constants.  
For the model of the force we use in this paper, 
which is equation \eqref{df_model}, if we
use the softening parameter proportional to the disk scale height,
$\epsilon = \epsilon_0 H(r)$, the constant $K$ is given by
\begin{equation}
 K = \dfrac{\pi G^2 M_{\rm p}\Sigma_0}{\epsilon_0 H_0 r_0^{-\alpha}},
\end{equation}
and $\alpha$ is given by
\begin{equation}
 \alpha = p + \dfrac{3-q}{2},
\end{equation}
where $H_0$ is the disk scale height at $r=r_0$.
Therefore, models with the same $K$ and $\alpha$ yield the same results.
In this section, we show the results on how $t_a$ and $t_e$ depend on
the disk profile $p$, or in more general terms, $\alpha$.  
We fix the value of surface density to be
$\Sigma_0/(M_{\ast}/r_0^2)=10^{-4}$, planet mass to be 
$M_{\rm p}/M_{\ast}=10^{-6}$, disk aspect ratio to be $H_0/r_0=0.05$,
and the softening parameter to be $\epsilon_0=0.5$.
The dependence of these parameters on the timescales $t_a$ and $t_e$ is
$t \propto M_{\rm p}^{-1} \Sigma_0^{-1} \epsilon_0 H_0$ as expected from
the form of $K$, as long as the force exerted by the disk is
sufficiently small compared to that from the central star. 
We use the disk model with $q=1$, which gives $\alpha = p+1$, but we
note again that the value of $\alpha$ is the key parameter 
which determines the values of $t_a$ and $t_e$.

Figure \ref{fig:tate_varalpha} shows $t_a$ and $t_e$ for
$\alpha=1,2,3,4$ ($p=0,1,2,3$, respectively for $q=1$).  
In Figure \ref{fig:tate_varalpha}, we fix $a/r_0=1$ and see how the
timescale behaves as we vary the eccentricity.  
We see that the behaviors of $t_a$ and $t_e$ strongly depend on the
imposed disk model.  The timescale of the evolution of semimajor axis
increases steadily as we increase eccentricity if $\alpha \leq 2$.  If,
on the other hand, $\alpha \geq 2$, the values of $t_a$ first increase
as we increase $e$, but decrease at large $e$.  The timescale of the
evolution of eccentricity always increases as we increase eccentricity
if $\alpha \leq 3$, but there is a maximum of $t_e$ at a certain value
of $e$ if $\alpha > 3$.  

As we decrease the values of $\alpha$, another interesting behavior
occurs for $t_a$.  In Figure \ref{fig:tate_varalpha2}, we show the
behavior of $t_a$ and $t_e$ for $\alpha=-1,0,1$ ($p=-2,-1,0$).  We see
that the timescale of the eccentricity evolution does not vary much
within this parameter range, 
while $t_a$ is negative for almost all the values 
of $e$ for $\alpha=0$ and $-1$.  The negative sign of $t_a$ indicates
that the direction of the semimajor axis evolution is outward.  

The reason why we obtain the outward semimajor axis evolution can be
qualitatively explained as follows.  If the index $p$ is negative, the
surface density increases as a function of radius.  As we have seen
before, the planet feels the tailwind at the aphelion and therefore is
exerted positive torque by the gas, 
while it feels headwind at the perihelion and is 
exerted negative torque.  Since the surface density is larger at
the aphelion than at the perihelion when $p$ is negative, 
the planet feels overall positive torque, and therefore the semimajor
axis increases.   

The disk with negative $p$ seems rather unrealistic.  However, the
surface density may increase as a function of radius {\it locally}.
The inner edge of the disk is one example where such local increase of 
the surface density is expected.  
If a planet with a finite eccentricity is 
placed in such a place, we expect that the semimajor axis can increase
as a result of the disk-planet interaction.  
Recently, \citet{ODI10} suggested that the planets trapped in mean
motion resonances can be stopped at the disk inner edge. 
This is because at the gap edge, 
 the planet feels the disk only at the place close to the aphelion and 
 therefore it is exerted the positive torque by the disk.  This positive
 torque is balanced by the negative torque exerted by the other planet
 in a mean motion resonance.

\subsection{Analytic Considerations for $e \sim 1$}

We have seen how the timescales of the semimajor axis evolution and
the eccentricity damping vary as we change the disk parameter.
Especially, we have seen a strong dependence of $t_a$ and $t_e$ on the
disk parameter $\alpha = p+(3-q)/2$ if the planet eccentricity is high.
In this section, we analytically investigate the behaviors of $t_a$ and
$t_e$ in the limit of $e \to 1$.  

In this section, we approximate $df/dt$ (see equation \eqref{Gauss_f})
by  
\begin{equation}
 \dfrac{df}{dt} = \dfrac{a^2 n\eta}{r^2},
\end{equation}
since the contribution from the perturbing force is small compared to
the force from the central star.  We then obtain
\begin{eqnarray}
 \dfrac{da}{df} &=& - K 
  \dfrac{2(1-e^2)^{2-\alpha}}{a^{2+\alpha}n^4}
  \nonumber \\
&&  \times 
 \dfrac{e^2 \sin^2 f + (1+e\cos f)^{3/2} (\sqrt{1+e \cos f}-1)}
  {(1+e\cos f)^{2-\alpha} 
  \left[ e^2 \sin^2 f + (1+\cos f)(2+e\cos f - 2\sqrt{1+e\cos f})
  \right]^{3/2}}
  \label{dadf_explicit}
\end{eqnarray}
and
\begin{eqnarray}
 \dfrac{de}{df} &=& - K 
  \dfrac{(1-e^2)^{3-\alpha}}{a^{3+\alpha}n^4}
  \dfrac{1}{(1+e^2+2e\cos f)(1+e\cos f)^{5/2-\alpha}}
  \nonumber \\
&&  \times 
 \Bigg\{
 \dfrac{\sin^2 f 
 \left[ 2e^2 (e+\cos f)\sqrt{1+e\cos f}+e(1-e^2) \right]}
 {\left[ e^2 \sin^2 f + (1+\cos f)(2+e\cos f - 2\sqrt{1+e\cos f})
  \right]^{3/2}}
 \nonumber \\ 
&& + \dfrac{2 (e+\cos f)(1+e \cos f)^2 (\sqrt{2+e \cos f}-1)}
  {\left[ e^2 \sin^2 f + (1+\cos f)(2+e\cos f - 2\sqrt{1+e\cos f})
  \right]^{3/2}}
  \Bigg\}
  \label{dedf_explicit}
\end{eqnarray}

From Figure \ref{fig:adotedot}, 
we see that the force exerted when the planet is near the aphelion 
and the perihelion is important in determining the orbital
evolution of the planet.  
We therefore look at the places close to the perihelion or aphelion.
Around these points, $\cos f$ and $\sin f$ are approximated by
\begin{equation}
 \cos f \sim \pm \left(1 + \dfrac{1}{2} \delta f^2 \right)
\end{equation}
\begin{equation}
 \sin f \sim \pm \delta f,
\end{equation}
where we write $f=\delta f$ near the perihelion and $f=\pi + \delta f$
near the aphelion, and the upper sign is for the perihelion and the
lower sign is for the aphelion.  
We expand equations \eqref{dadf_explicit} and
\eqref{dedf_explicit} up to the second order of $\delta f$.
After the expansion with respect to $\delta f$, we take the limit of
$e \to 1$.  We define
\begin{equation}
 \varepsilon = 1-e
\end{equation}
and take the lowest order of $\varepsilon$.  

The calculations are tedious but straightforward.  In the close vicinity
of the perihelion, we obtain
\begin{equation}
 \dfrac{da}{df} \sim - K
  \dfrac{\varepsilon^{2-\alpha}}{a^{2+\alpha}n^4}
  \dfrac{2(\sqrt{2}-1)}{(3-2\sqrt{2})^{3/2}}
  \dfrac{1+\delta f^2 (2-\sqrt{2})/8\sqrt{2}}
  {(1-\delta f^2/4)^{2-\alpha}
  \{1+\delta f^2 3(\sqrt{2}-1)/4(3-2\sqrt{2})\}^{3/2}}
  \label{peridadf_raw}
\end{equation}
and
\begin{equation}
 \dfrac{de}{df} \sim K 
  \dfrac{\varepsilon^{3-\alpha}}{a^{3+\alpha}n^4}
  \dfrac{(\sqrt{2}-1)(1-\delta f^2 (5+6\sqrt{2})/4\sqrt{2})}
  {\sqrt{2}(3-2\sqrt{2})(1-\delta f^2/4)^{7/2-\alpha}
  (1+\delta f^2 3(\sqrt{2}-1)/4(3-2\sqrt{2})}.
  \label{peridedf_raw}
\end{equation}
We then integrate over the orbit.  We denote the integral by,
\begin{equation}
 \overline{\dfrac{da}{df}} = \int_0^{2\pi} \dfrac{da}{df} df.
\end{equation}
If we integrate over $f$, the terms in equations \eqref{peridadf_raw}
and \eqref{peridedf_raw} that contain $\delta f$ result in a 
numerical factor of the order of unity.  Therefore, close to the
perihelion, $da/df$ and $de/df$ averaged over one orbital period are  
\begin{equation}
 \overline{\dfrac{da}{df}} \propto a^{4-\alpha} \varepsilon^{2-\alpha} 
\end{equation}
and 
\begin{equation}
 \overline{\dfrac{de}{df}} \propto a^{3-\alpha} \varepsilon^{3-\alpha}.
\end{equation}

We now turn the attention to the aphelion.  
In the close vicinity of the aphelion, we obtain
\begin{equation}
 \dfrac{da}{df} \sim 2^{3-\alpha} K
  \dfrac{1}{a^{2+\alpha}n^4}
  \dfrac{1-\delta f^2/\varepsilon^{3/2}}
  {(1+\delta f^2/2\varepsilon)^{2-\alpha}
  (1+3\delta f^2/2\varepsilon)^{3/2}}
\end{equation}
and 
\begin{equation}
 \dfrac{de}{df} \sim -2^{3-\alpha} K 
  \dfrac{1}{a^{3+\alpha}n^4 \varepsilon^{1/2}}
  \dfrac{1+\delta f^2/\varepsilon^{3/2}}
  {(1+\delta f^2/2\varepsilon)^{5/2-\alpha}
  (1+\delta f^2/\varepsilon^2)(1+3\delta f^2/2\varepsilon)^{3/2}}. 
\end{equation}
We now integrate over $f$ near the aphelion.  
The integration for $da/df$ is rather complicated.  We have
\begin{equation}
 \mathcal{I}_a = \int d \delta f 
  \dfrac{1-\delta f^2/\varepsilon^{3/2}}
  {(1+\delta f^2/2\varepsilon)^{2-\alpha}
  (1+3\delta f^2/2\varepsilon)^{3/2}}
\end{equation}
and changing the integration variable to $X=\delta f/\varepsilon^{1/2}$, 
\begin{equation}
 \mathcal{I}_a = \varepsilon^{1/2} \int dX 
  \dfrac{1-X^2/\varepsilon^{1/2}}{(1+X^2/2)^{2-\alpha} (1+3 X^2
  /2)^{3/2}}.  
\end{equation}
If $\varepsilon$ is sufficiently small, the numerator is dominated by
the second term, and therefore the resulting $\mathcal{I}_a$ is negative 
and $\mathcal{I}_a \propto \varepsilon^0$.  If $\varepsilon$ is not very
small, the numerator is dominated by the first term.  Therefore the
resulting integral is positive and
$\mathcal{I}_a \propto\varepsilon^{1/2}$.  Therefore, the contribution of
the aphelion to $\overline{da/df}$ is
\begin{eqnarray}
 \overline{\dfrac{da}{df}}  \propto \left\{
  \begin{array}{cc}
   a^{4-\alpha} & \mathrm{negative \ for \ small \ \varepsilon} \\
   a^{4-\alpha} \varepsilon^{1/2} 
    & \mathrm{positive \ for \ intermediate \ \varepsilon}
  \end{array}
  \right.
\end{eqnarray}
The integration for $de/df$ is more straightforward.  We have
\begin{equation}
 \mathcal{I}_e = \int d \delta f 
  \dfrac{1+\delta f^2/\varepsilon^{3/2}}
  {(1+\delta f^2/2\varepsilon)^{5/2-\alpha}
  (1+\delta f^2/\varepsilon^2)(1+3\delta f^2/2\varepsilon)^{3/2}}
\end{equation}
and changing the variable to $X=\delta f/\varepsilon$, 
\begin{equation}
 \mathcal{I}_e = \varepsilon \int dX
  \dfrac{1+\varepsilon^{1/2}X^2}
  {(1+X^2)(1+\varepsilon X^2/2)^{5/2-\alpha} 
  (1+3\varepsilon X^2/2)^{3/2}} .
\end{equation}
The terms proportional to $\varepsilon^{1/2}$ and $\varepsilon$ are
safely neglected, and therefore we have 
$\mathcal{I}_e \propto \varepsilon$.  Therefore, for the contribution
from the aphelion to $\overline{de/df}$, we have
\begin{equation}
 \overline{\dfrac{de}{df}} \propto a^{3-\alpha}\varepsilon^{1/2}.
\end{equation}

It still remains to be determined whether contributions from the
perihelion or from the aphelion dominate.   
The relative importance of the contributions from the aphelion and
the perihelion depends on the disk model, 
and we numerically see which contribution is more important as we change
$\alpha$.   
In Figure \ref{fig:dade_ave}, 
we plot the values of $-\overline{da/df}$ and
$-\overline{de/df}$ for various disk models.  
We see that the contribution from the perihelion dominates when
$\alpha \gtrsim 2$ (corresponding to $p \gtrsim 1$ if $q=1$), 
and the expected behaviors of 
$\overline{da/df}\propto\varepsilon^{2-\alpha}$
and
$\overline{de/df}\propto\varepsilon^{3-\alpha}$
are observed.  
If $\alpha \lesssim 2$, the contribution from
the aphelion comes into play, and we see $\overline{de/df}$ is roughly
proportional to $\varepsilon^{1/2}$.  We note that for the model
with $\alpha=-1$ and $\alpha=0$, the direction of semimajor axis
evolution is outward for the 
moderate values of eccentricity ($\varepsilon>0.09$ for $\alpha=0$ model
and $\varepsilon>0.05$ for $\alpha=-1$), as we have already
seen in the previous section.  
The inversion of the sign of $\overline{da/df}$ 
is also expected from the above analytic discussions at the aphelion.  
We have also checked that 
the values of $\overline{da/df}$ are always positive if
we have a large negative value of $p$, when the interaction is
completely dominated by the contribution from the aphelion. 

We summarize the dependence of the semimajor axis evolution timescale
and eccentricity damping timescale on the disk parameters in the case of
highly eccentric planet.   
Since $\overline{da/dt}$ and $\overline{da/df}$ are related by 
\begin{equation}
 \overline{\dfrac{da}{dt}} = \dfrac{1}{T_{\rm orb}}
  \overline{\dfrac{da}{df}},
\end{equation}
$t_a \propto a^{5/2} \overline{da/df}^{-1}$.  In the same way, it is
possible to show that 
$
t_e \propto e a^{3/2} \overline{de/df}^{-1} 
\propto a^{3/2} \overline{de/df}^{-1}
$
, where we have used 
$e\sim 1$.  
In the case of $\alpha \gtrsim 2$, 
the contribution from the perihelion dominates and therefore,
\begin{equation}
 t_a \propto a^{\alpha-3/2} \varepsilon^{\alpha-2}
  \label{ta_anal1}
\end{equation}
\begin{equation}
 t_e \propto a^{\alpha-3/2} \varepsilon^{\alpha-3}.
  \label{te_anal1}
\end{equation}
The evolution of the semimajor axis is always inward and the
eccentricity always damps.
If $\alpha \lesssim 2$, we take into account the contribution from the
aphelion, 
\begin{eqnarray}
 t_a  \propto \left\{
  \begin{array}{cc}
   a^{\alpha-3/2} & \mathrm{inward \ for \ small \ \varepsilon} \\
   a^{\alpha-3/2} \varepsilon^{-1/2} 
    & \mathrm{outward \ for \ intermediate \ \varepsilon}
  \end{array}
  \right.
 \label{ta_anal2}
\end{eqnarray}
and
\begin{equation}
 t_e \propto a^{\alpha-3/2}\varepsilon^{-1/2}.
  \label{te_anal2}
\end{equation}
It is noted that the outward evolution of semimajor axis occurs only
when $\alpha \lesssim 0$ and eccentricity always damps.  The range of
$\varepsilon$ where the planet experiences the outward semimajor axis
evolution increases as the value of $\alpha$ is decreased.  
In the particular disk model with $q=1$, $\alpha$ and $p$ are related
by $\alpha = p+1$.

We have shown that there is a critical value for the power of the
surface density, which determines the behavior of semimajor and
eccentricity evolutions for a highly eccentric planet ($e \sim 1$).  
Let us consider the model with $q=1$.  
If $p>2$, both eccentricity and semimajor axis evolution
timescale becomes small as we increase the planet's eccentricity.  If
$1<p<2$, semimajor axis evolution timescale becomes shorter for higher
eccentricity, while eccentricity evolution timescale becomes longer.
For $p<1$, both evolution timescales become longer as we increase the
eccentricity of the planet.  Further complication arises for the
semimajor axis evolution for $p \lesssim 0$, 
where the contribution from the aphelion comes into play to invert the
direction of the evolution. 

However, we note that these critical values of $p$ should not be
overstated.  
In the model described above, the critical parameter that determines
the behavior of the orbital evolution timescales is $\alpha=p+(3-q)/2$.   
The appearance of this single parameter $\alpha$ 
is partly due to the prescription of 
the cutoff of the gravitational force.  Note
that the dynamical friction force we have used depends on the cutoff
scale $\epsilon$ and we have used the model in which $\epsilon$ depends 
linearly on the disk scale height.  The appearance of $q$ in the
parameter $\alpha$ depends on this simplified prescription of the
cutoff.  
We also note that the dependence of the timescales on the softening
length is different from \citet{PL00}, which is due to the difference in
the formulation.

Therefore, we consider that the critical values of $p$ described
above only have qualitative meanings.  
Yet, we expect that the model describes the qualitative behavior of the 
semimajor axis and eccentricity evolution of highly eccentric planets.
The prediction is readily compared with two-dimensional numerical
calculations, although a very large simulation box may be necessary.  
In order to get rid of the dependence on the softening parameter
$\epsilon$, it is necessary to perform three-dimensional analyses.

\subsection{Validity of the Model}

In this section, we briefly discuss the applicability of our model.  
The most important assumption we have used in the model is the use of
the supersonic dynamical friction formula \eqref{df_supersonic_limit},
which is derived from linear perturbation analyses.  
The use of this formula assumes (1) the relative velocity is supersonic,
(2) the flow in the vicinity of the planet is homogeneous, and (3) the
steady state is reached instantaneously.  We shall now discuss each
assumption separately.  We also discuss how non-linear effects on the
dynamical friction force may change our results.

The first assumption is justified if we consider a planet that has high 
enough eccentricity, as indicated in Figure \ref{fig:relv_vare}.
Typically, when $e \gg H/r$, we can safely assume that the flow is
supersonic.  

For the second assumption, let us consider which scale of the
perturbation contributes to the force.  
In the discussion of Section \ref{sec:df}, we
have seen that the perturbation of 
all the scales larger than the cutoff 
length equally contributes to the dynamical friction force.  
Let us consider a planet embedded in a disk with Keplerian
rotation, and the planet is located at the perihelion or aphelion.  
Then, there is a location in the disk where 
the relative velocity between the gas and the planet becomes zero.  
We call this radius an ``instantaneous corotation radius'', $r_C$, and
is given by, for the perihelion, 
\begin{equation}
 \dfrac{GM_{\ast}}{r_C} = \dfrac{GM_{\ast}}{a} \dfrac{1+e}{1-e}.
\end{equation}
A similar equation can be derived for the aphelion.  

The assumption of the homogeneous flow breaks down if the length scale
comparable to the distance between the planet's location and the
instantaneous corotation radius is important in determining the force
acting on the planet.  
If the planet's eccentricity is large, the distance between the
instantaneous corotation radius $r_C$ and the planet's location is of
the order of the scale of the disk itself ($|r_C - a| \sim a$), which is
much larger than the cutoff scale, which is 
comparable with the disk scale height.  Therefore, in this case, we
expect that the assumption of the homogeneous flow is justified.  
If the planet's eccentricity is small, the instantaneous corotation radius
comes very close to the planet's orbital radius ($r_C \sim a$), 
and the assumption of 
homogeneous medium becomes worse.  Typically, we expect that this
assumption breaks down if the planet's 
eccentricity is smaller than the disk aspect ratio, and in such a small 
eccentricity case, it is necessary to fully take into account the
effects of shear around the planet's location.

Regarding the third assumption, the time taken to reach the steady state
may be estimated as follows.  The length scale that is important in
determining the force is of the order of the disk scale height, and the
minimum speed of the propagation of the information of the perturbing
potential may be the sound speed.  This indicates that the time taken to
develop the steady state is of the order of the Kepler timescale.
However, since the relative velocity is supersonic, the time taken to
develop the steady state is less than that, since the background flow
will carry the information of the perturber.  In any case, this
assumption is only marginally satisfied.  

As discussed before, it is known from numerical simulations 
that the time when the maxima/minima of the torque occurs lags the time
of the maxima/minima of the distance between the planet and the
central star \citep[e.g.,][]{CDKN07}.  Our model does not give a
precise prescription for this effects, and the instantaneous torque or
power profiles are different from what we expect from the numerical
simulations.  However, if we take the average over one orbital period,
our model is in good agreement with the previous calculations in
the parameter range where both our model and the previous 
calculations are expected to give reasonable results.   

In short, we expect that our formulation is applicable to planets whose
eccentricity is larger than the disk aspect ratio, in particular when
one takes the average over one orbital period.

Finally, we make a comment on the use of the dynamical friction formula
derived by the linear perturbation analysis.  \citet{KK09} investigated
the dynamical friction using the non-linear numerical simulations.
They considered the case where the particle is embedded in a
homogeneous, three-dimensional gas flow, and calculated the
axisymmetric pattern of the gas flow induced by the particle's gravity.
They found that 
there is a deviation from the linear theory depending on the
Mach number $M_0$ of the flow and the non-linear parameter 
\begin{equation}
 \mathcal{A} = \dfrac{GM}{c^2 r_S},
\end{equation}
where $r_S$ is the (effective) radius of the body.  The force exerted on
the particle is deviated from the linear results by a factor of
$(\eta/2)^{-0.45}$, where $\eta=\mathcal{A}/(M_0^2-1)$.  In the case of
a planet embedded in a protoplanetary disk, $\mathcal{A}$ is of the
order of $10-100$, and therefore, non-linearity can be important.  The
correction factor, however, is the quantity of the order of unity,
(since Mach number is of the order of 5-10) and
therefore, our results may give a reasonable estimate of the timescale
of the evolution of the orbital parameter.  One possible effect of such
non-linearity on our results is that the dependence of the timescales
$t_a$ and $t_e$ on the orbital parameters of the planet can be
different, since the force depends on the velocity of the perturbing
potential in a different way.

\section{Discussion: Possible Applications to Other Studies}
\label{sec:discuss}

In this section, we discuss some possible applications of our results to
other studies and the implications to the planet formation theory.  
We first show the fitting formula for $t_a$ and $t_e$ derived from this
study, which can be especially useful for population synthesis models
such as \citet{IL08}.  
We then discuss some implications from this study to recent planet
formation scenarios.

\subsection{Fitting Formula for Timescale of Semimajor Axis and
  Eccentricity Evolution}

In this section, we derive the fitting formula for $t_a$ and $t_e$.
As seen before, within our framework,  
$t_a, t_e \propto M_p^{-1}\Sigma_0^{-1} \epsilon_0 H_0$ and the only disk
parameter that controls the timescale is $\alpha = p + (3-q)/2$.
Therefore, we fit the values of $t_a$ and $t_e$ 
as a function of the semimajor axis and the eccentricity  
for a given value of $\alpha$ with $1 \leq \alpha \leq 4$.  
For the data for fitting, we use the values 
obtained for $0.3 \leq a/r_0 \leq 10$ and 
$0.1 \leq e \leq 0.99$.  

Motivated by the results of the analytic considerations for $e \sim 1$,
we use the fitting formula of the form: 
\begin{equation}
 t_a = C_a(a) e^{\zeta_a(a)} (1-e^{\lambda_a(a)})^{\eta_a(a)}
  \left( \dfrac{M_p/M_{\ast}}{10^{-6}}  \right)^{-1}
  \left( \dfrac{\Sigma_0/M_{\ast}r_0^{-2}}{10^{-4}}  \right)^{-1}
  \left( \dfrac{h_0}{0.05}  \right)
  \left( \dfrac{\epsilon_0}{0.5}  \right)
  \label{fiteq_ta}
\end{equation}
Here, $C_a$, $\zeta_a$, $\lambda_a$ and $\eta_a$ are the function of $a$
of the form
\begin{eqnarray}
&& C_a(a) = C_{a0} \left( \dfrac{a}{r_0} \right)^{C_{ap}} \\
&& \zeta_a(a) = \zeta_{a0} \left( \dfrac{a}{r_0} \right)^{\zeta_{ap}} \\
&& \lambda_a(a) = \lambda_{a0} 
 \left( \dfrac{a}{r_0} \right)^{\lambda_{ap}} \\
&& \eta_a(a) = \eta_{a0} \left( \dfrac{a}{r_0} \right)^{\eta_{ap}}.
\end{eqnarray}
The same form of the fitting function is used for $t_e$,
\begin{equation}
 t_e = C_e(a) e^{\zeta_e(a)} (1-e^{\lambda_e(a)})^{\eta_e(a)}
  \left( \dfrac{M_p/M_{\ast}}{10^{-6}}  \right)^{-1}
  \left( \dfrac{\Sigma_0/M_{\ast}r_0^{-2}}{10^{-4}}  \right)^{-1}
  \left( \dfrac{h_0}{0.05}  \right)
  \left( \dfrac{\epsilon_0}{0.5}  \right), 
  \label{fiteq_te}
\end{equation}
with
\begin{eqnarray}
&& C_e(a) = C_{e0} \left( \dfrac{a}{r_0} \right)^{C_{ep}} \\
&& \zeta_e(a) = \zeta_{e0} \left( \dfrac{a}{r_0} \right)^{\zeta_{ep}} \\
&& \lambda_e(a) = \lambda_{e0} 
 \left( \dfrac{a}{r_0} \right)^{\lambda_{ep}} \\
&& \eta_e(a) = \eta_{e0} \left( \dfrac{a}{r_0} \right)^{\eta_{ep}}.
\end{eqnarray}
We note that in the set of the fitting parameters, $C_{a0}$ and $C_{e0}$ 
have the dimension of time, while others are dimensionless.  

In Tables \ref{table:fitta} and \ref{table:fitte}, we show the results
of fitting for $t_a$ and $t_e$, respectively.  
For $C_{a0}$ and $C_{e0}$, we give the values in the units of year with
$M_{\ast}=M_{\odot}$ and $r_0 = 1\mathrm{AU}$.  In the 
last column of the tables, we show the maximum values of the error of the
fitting function, which is calculated as
$|t_{\rm calc}-t_{\rm fit}|/t_{\rm calc}$, 
where $t_{\rm calc}$ is the value of $t_a$ or $t_e$ 
calculated in the model presented in this paper
and $t_{\rm fit}$ is the value of the fitting function.  Our fitting
formula successfully reproduce the values of $t_a$ and $t_e$ within
$10-20\%$ error.  We also note that $a^{\alpha-3/2}$ dependence that
appears in equations \eqref{ta_anal1}-\eqref{te_anal2} can be clearly
seen in the fitting parameters $C_{ap}$ and $C_{ep}$. 

Here, we note that the caution must be payed in using the fitting
formulae.  In obtaining the fitting parameters, we have used the values
with $0.3 \leq a/r_0 \leq 10$ and $0.1 \leq e \leq 0.99$.  
Especially, our formulation is not applicable to the case of a 
low-eccentricity planet (typically $e \lesssim H/r$).  
It is necessary to combine the results of
this paper and such formulae given by \citet{TW04} which are applicable
to the case of low eccentricity.  

It is also noted that the fitting formulae are given in the case of 
$1 \leq \alpha \leq 4$.  Especially, our form of fitting
functions \eqref{fiteq_ta} and \eqref{fiteq_te} fails to describe the
change of the sign of $t_a$ that happens in the case of $p \lesssim -1$
(see Figure \ref{fig:dade_ave}).  
Although we believe that this range of $\alpha$
covers most of the typical protoplanetary disk parameters, it is
necessary to construct the fitting formula if one wants to apply for
protoplanetary disk parameters out of the range of this fitting.  
Even in this case, however, we expect that the model 
described in this paper, which uses the dynamical friction force to
calculate the evolution of orbital parameters, is still applicable.

\subsection{Implication to Planet Formation Theory}

We have developed a model for the disk-planet interaction that
can be used especially for highly eccentric planets.  
We have found that in the case of a highly eccentric planet, the
disk-planet interaction is essentially described by dynamical friction. 
We have derived how the evolution timescales of the planet's semimajor
axis and eccentricity depend on disk parameters within our models.  In
this section, we discuss the possible implications to the planet
formation theory.  

Recently, \citet{MIM11} have suggested that a disk surrounding a newly
born protostar is gravitationally unstable, and therefore it may be
possible to form a planet by gravitational instability 
\citep[see also][]{IMM10}.  
The planets born in such a way are expected to have high eccentricity.  
Also, it is possible to form a planet at a large distance from the
central star.   

We briefly discuss the evolution timescale of such planets.  
From the fitting formula of the evolution timescales, it is possible to
show that the orbital evolution timescales of the newly-born
Jupiter-mass planets through the gravitational instability can be very
rapid, since the disk is very massive compared to the central star.
However, after the accretion phase onto the central star, the stellar
mass becomes much larger than the disk mass.  If there is a low-mass
planet (e.g., sub-Jupiter mass) that has 
survived in the early phase and reasonable amount of eccentricity is
maintained, the orbital evolution timescale can be of the comparable
order of magnitude with the observed disk lifetime \citep{HLL01}.  
Such planets may be found by the direct imaging observations (e.g.,
\citet{M08}), although small mass planets may be very cold and the
observations may be difficult.  
Nevertheless, a highly eccentric planet at a 
large orbital separation can be a good observational signature to test
the theory of planet formation and disk-planet interaction.

Another possible mechanism to produce a planet in a highly eccentric
orbit is the planet-planet scattering (e.g., \citet{CWB96,MW02}).  The
scattering between planets can naturally result in the highly eccentric
planets.  From the calculations given in this paper, 
the evolution timescale of the orbital parameters can be very long even 
if the planet's orbital plane coincides with the gas disk plane.  
If the planet's orbital plane is not aligned with the plane of the disk,
we expect that the evolution timescale becomes much larger, since the
planet interacts with the disk only at a very small part of the orbit.  
The calculations given in this paper sets the lower limit of the
timescales of the orbital evolution for such planets.  
It is therefore possible that the scattered planets remain in a highly
eccentric orbit, even if gas disk remains and the tidal interaction with
the central star is not effective.  

Another interesting application of our model is the recently
suggested ``eccentricity trap'' mechanism \citep{ODI10}.  They have
suggested that if there is an inner cavity in a protoplanetary disk, it
may be possible to maintain a planet at the inner edge of the disk.  It
is due to the positive torque acting on the planet at the aphelion,
which stops the planet to migrate inward.  
Although our@formulation, which assumes a smooth disk profile, is not
directly applicable to the disk with a sharp inner edge, we have
already seen that the evolution of semimajor axis can be outward if
there is a steep surface density gradient.  It is an interesting
extension to apply the method of dynamical friction to the disk with a
cavity.

\section{Summary}
\label{sec:summary}

In this paper, we have presented a model of the
disk-planet interaction, which makes use of the dynamical friction 
force.  This method can be especially useful for a highly
eccentric planet.   

We have first derived the dynamical friction
formula in a slab media using different methods from \citet{Ost99}, and
calculated the dynamical friction force as a function of viscosity and
Mach number.  The integral that leads to the dynamical friction force is
given by equation \eqref{Force_int}, and some asymptotic expressions are
given by equations \eqref{df_subsonic}, \eqref{df_subsonic_limit},
\eqref{df_supersonic}, and \eqref{df_supersonic_limit}.

We have then applied the dynamical friction formula to find the
evolution of orbital parameters of an eccentric planet embedded in a
disk.  The results agree well with the previous calculations given by
\citet{PL00} for moderate eccentricity 
($e\gtrsim H/r$), and we expect that it is
possible to use our model to much higher eccentricity.  
In this sense, this paper provides a complement to the previous study.
We have seen that the evolution of the orbital
parameters depends on the disk structure, and calculated some critical
values for $p$, the power of the surface density profile, at which the
behaviors of the timescales of orbital evolutions are qualitatively
altered.  

Our model presented in the paper is restricted to a planet with a
high eccentricity, typically $e \gtrsim H/r$.  
It is also necessary that the disk profile must be smooth in the
vicinity of the planet, in order to justify the use of dynamical
friction formula, which is obtained under the assumptions of a
homogeneous slab.

\acknowledgments
T.M. thanks Cl\'{e}ment Baruteau for useful discussions.  
This work was supported in part by the Grants-in-Aid for Scientific
Research 22$\cdot$2942 (T.M.) and 20540232 (T.T.) 
of the Ministry of Education, Culture, Sports, Science, and Technology
(MEXT) of Japan.

\appendix

\section{Time Dependent Analysis of Dynamical Friction Force}
\label{app:timedep}

In this appendix, we perform a time-dependent linear analysis for the 
gravitational interaction between a slab and an embedded particle.
We do not perform the analysis in full details.  The objective of
this section is to show that it is sufficient to perform the
steady-state analysis in order to obtain the dynamical friction force.  

We start with equation \eqref{wave_lin}, and perform Fourier transform
in the spatial direction via equations \eqref{FT_def} and
\eqref{invFT_def}.  We then obtain an ordinary differential equation in
time as
\begin{equation}
 \dfrac{d^2 \tilde{\alpha}}{dt^2} 
  + \left( 2ik_y v_0 + \nu k^2 \right) 
  \dfrac{d\tilde{\alpha}}{dt}
  - \left( c^2 k^2 -k_y^2 v_0^2 
     + i k^2 k_y v_0 \nu   \right) \tilde{\alpha}
  = -k^2 \tilde{\psi}.
  \label{timedep_eq}
\end{equation}
We consider that at time $t=0$, the particle is suddenly switched on,
\begin{eqnarray}
 \tilde{\psi} (t,k) = \left\{
  \begin{array}{cc}
   0 & (t<0) \\
   -\dfrac{GM}{k}e^{-k\epsilon} & (t>0)
  \end{array}
  \right.
\end{eqnarray}
We define $p,q,S$ (they should not be confused with $p$ and $q$ in the
main text) as
\begin{equation}
 p \equiv \nu k^2 + 2i k_y v_0,
\end{equation}
\begin{equation}
 q \equiv c^2 k^2 - v_0^2 k_y^2 + i k^2 k_y v_0 \nu,
\end{equation}
and
\begin{equation}
 S \equiv -k^2 \tilde{\psi}
\end{equation}
so equation \eqref{timedep_eq} can be written in a concise way
\begin{equation}
 \dfrac{d^2 \alpha}{dt^2} + p\dfrac{d\alpha}{dt} 
  + q \alpha = S,
  \label{eq_simple}
\end{equation}
where we have also removed the tilde sign for convenience.  This
equation should not be confused with that in the real space.  

If we define the function $g$ by
\begin{equation}
 \alpha(t) = e^{-pt/2} g(t),
\end{equation}
equation \eqref{eq_simple} becomes 
\begin{equation}
 \dfrac{d^2 g}{dt^2} 
  + \left( q - \dfrac{p^2}{4} \right) g
  = \mathcal{S},
\end{equation}
where $\mathcal{S}\equiv e^{pt/2} S$.

The boundary condition is that there is no perturbation at $t<0$.  The
solution that satisfies this condition is readily obtained as
\begin{equation}
 g(t) = \dfrac{1}{\sqrt{p^2-4q}} 
  \left[
   g_{+}(t) \int_0^t \mathcal{S}(\tau) g_{-}(\tau) d\tau
   - g_{-}(t) \int_0^t \mathcal{S}(\tau) g_{+}(\tau) d\tau
  \right],
  \label{g_sol}
\end{equation}
where
\begin{equation}
 g_{\pm}(t) = \exp\left[\pm t \sqrt{\dfrac{p^2}{4} - q}\right]
\end{equation}
are the homogeneous solutions.  
From equation \eqref{g_sol}, we can write down the solution of 
$\alpha(t)$ as
\begin{eqnarray}
 \alpha(t) &=& \dfrac{S}{q}   
  \Bigg\{
   1  
 - \dfrac{p/2 + \sqrt{p^2/4-q}}{\sqrt{p^2-4q}}
   \exp\left[-\left(\dfrac{p}{2} 
	       -\sqrt{\dfrac{p^2}{4}-q}\right)t \right]
 \nonumber  \\
&&   + \dfrac{p/2 - \sqrt{p^2/4-q}}{\sqrt{p^2-4q}}
   \exp\left[-\left(\dfrac{p}{2} 
	       +\sqrt{\dfrac{p^2}{4}-q}\right)t \right]
  \Bigg\}
  \label{alpha_sol}
\end{eqnarray}

We first note that the arguments in the exponential in equation
\eqref{alpha_sol} is given by
\begin{equation}
 \dfrac{p}{2} \pm \sqrt{\dfrac{p^2}{4}-q} 
  = \dfrac{1}{2} \nu k^2 + ik_y v_0
  \pm \sqrt{\dfrac{\nu^2(k_x^2+k_y^2)^2}{4} - c^2 k^2}
\end{equation} 
and therefore, the real part is always positive in the presence of
viscosity.  Therefore, in the limit of $t\to \infty$, the contribution
from the last two terms in equation \eqref{alpha_sol} vanishes if there
is a viscosity, and the steady state solution
\begin{equation}
 \alpha = \dfrac{S}{q}
\end{equation}
is reached.  

We now discuss the inviscid case, where the last two terms in equation
\eqref{alpha_sol} oscillates.  We show even in this case, the
contribution to the dynamical friction force coming from these
time-dependent terms vanishes as $t^{-1}$.    

The expression of dynamical friction force \eqref{force_Four} still
holds in the time-dependent analysis.  Therefore, we investigate the
integral
\begin{equation}
 I_t \equiv \int_0^{\infty} dk_y \int_{-\infty}^{\infty} dk_x
  k_y \dfrac{e^{-k\epsilon}}{k} \alpha,
  \label{Fint}
\end{equation}
where the factor $e^{-k\epsilon}/k$ comes from the Fourier transform of
the gravitational potential.  In the case of an inviscid gas, the
surface density perturbation is given by
\begin{eqnarray}
 \alpha(t) &=& \dfrac{GMe^{-k\epsilon}}{c^2 k (1-M_0^2k_y^2/k^2)}   
  \nonumber \\
&& \times
  \Bigg\{
   1  
 - \dfrac{1}{2}
 \left( \exp\left[ick(1-M_0\dfrac{k_y}{k})t\right]
  + \exp\left[-ick(1+M_0\dfrac{k_y}{k})t\right]
 \right)
 \nonumber \\
&& -\dfrac{M_0}{2}\dfrac{k_y}{k}
 \left( \exp\left[ick(1-M_0\dfrac{k_y}{k})t\right]
  - \exp\left[-ick(1+M_0\dfrac{k_y}{k})t\right]
 \right)
  \Bigg\}
  \label{alpha_sol_inviscid}
\end{eqnarray}
We consider the terms in the integral \eqref{Fint} involving the
time-dependence, which are second and the third terms in the curly
bracket of equation \eqref{alpha_sol_inviscid}.
Changing the variables as done in equations \eqref{kx_trans}
 and \eqref{ky_trans}, we obtain the terms that are proportional to 
\begin{equation}
 I_n = \int_0^{\infty}dk \int_0^{\pi} d\theta
  \dfrac{\sin^{n}\theta e^{-2k\epsilon}}{1-M_0^2 \sin^2 \theta}
  \exp\left[\pm ickt (1\mp M_0 \sin\theta) \right],
\end{equation}
where $n=1$ for the second term and $n=2$ for the third term.  We
further change the variable from $k$ to $\psi=ckt$, and perform the
integration over $\psi$.  We now have
\begin{equation}
 \mathrm{Im}\left(I_n\right) 
  = \pm \dfrac{1}{ct} \int_0^{\pi} d\theta
  \dfrac{\sin^{n}\theta}
  {\left\{(2\epsilon/ct)^2 + (1 \mp M_0\sin\theta)^2 \right\}
  \left(1\pm M_0 \sin\theta\right)},
  \label{df_timedep}
\end{equation}
where we have taken the imaginary part which contributes to the
dynamical friction force.  The integral involved in equation
\eqref{df_timedep} is finite for all the values of $t>0$, 
and asymptotes to a finite value as $t\to\infty$.  Therefore, 
the contribution to the force arising from the time-dependent terms
decays as $t^{-1}$, which justifies the steady-state assumption given in 
the main text.

\section{The Exact Expression of $I_{\theta}$}
\label{app:intI}

The integral given by equation \eqref{intI_expression} can be performed
analytically.  We make use of {\it Mathematica} software to calculate
the integral and we show only the results in this section.  
The integral is given by
\begin{equation}
 I_{\theta}(p,q) = \dfrac{A}{B},
\end{equation}
where the numerator is 
\begin{eqnarray}
 \dfrac{A}{2} &=& \sqrt{2} p^2 \left(-\sqrt{q} \sqrt{q-4 p}-2
			   p+q+2\right)  \nonumber \\
 &&  \times 
  \log \left(\frac{2 p^2-\sqrt{q} \sqrt{q-4 p}-2 p+q}{\sqrt{2 p^3+p^2
   \left(\sqrt{q} \sqrt{q-4 p}-q-2\right)+q^{3/2} \sqrt{q-4 p}+p
   \left(4 q-2 \sqrt{q} \sqrt{q-4 p}\right)-q^2}}\right) \nonumber \\
 && +\sqrt{2} p^2
   \left(\sqrt{q} \sqrt{q-4 p}+2 p-q-2\right) \nonumber \\
&& \times
 \log \left(\frac{-2
   p^2+\sqrt{q} \sqrt{q-4 p}+2 p-q}{\sqrt{2 p^3+p^2 \left(\sqrt{q}
   \sqrt{q-4 p}-q-2\right)+q^{3/2} \sqrt{q-4 p}+p \left(4 q-2 \sqrt{q}
   \sqrt{q-4 p}\right)-q^2}}\right) \nonumber \\
 && + \pi  \left(\sqrt{q} \sqrt{q-4 p}-2
   p+q\right)  \\
&& \times  \sqrt{2 p^3+p^2 \left(\sqrt{q} \sqrt{q-4
   p}-q-2\right)+q^{3/2} \sqrt{q-4 p}+p \left(4 q-2 \sqrt{q} \sqrt{q-4
   p}\right)-q^2} \nonumber \\ 
 && \times \sqrt{\frac{-2 p^4+4 p^3-2 p^2 \left(\sqrt{q}
   \sqrt{q-4 p}+q+1\right)-q^{3/2} \left(\sqrt{q-4 p}+\sqrt{q}\right)+2
   p \left(\sqrt{q} \sqrt{q-4 p}+2 q\right)}{2 p^3-p^2 \left(\sqrt{q}
   \sqrt{q-4 p}+q+2\right)-q^{3/2} \left(\sqrt{q-4 p}+\sqrt{q}\right)+2
   p \left(\sqrt{q} \sqrt{q-4 p}+2 q\right)}} 
   \nonumber
\end{eqnarray}
and the denominator is 
\begin{eqnarray}
B &=& \sqrt{q}
   \sqrt{q-4 p} \left(2 p^2+\sqrt{q} \sqrt{q-4 p}-2 p+q\right) 
   \nonumber \\ 
 &&  \times  \Bigg( 2
   p^3+p^2 \left(\sqrt{q} \sqrt{q-4 p}-q-2\right) \nonumber \\
 && +q^{3/2} \sqrt{q-4
   p}+p \left(4 q-2 \sqrt{q} \sqrt{q-4 p}\right)-q^2 \Bigg)^{1/2} .
\end{eqnarray}

\clearpage

\begin{figure}
 \plottwo{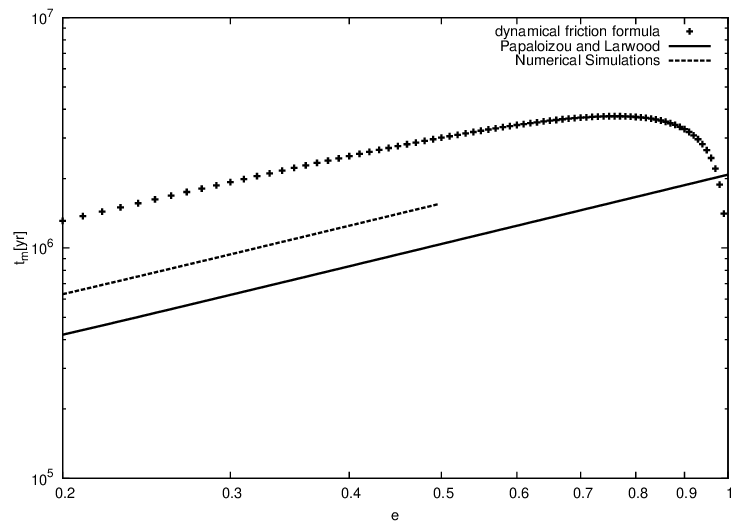}{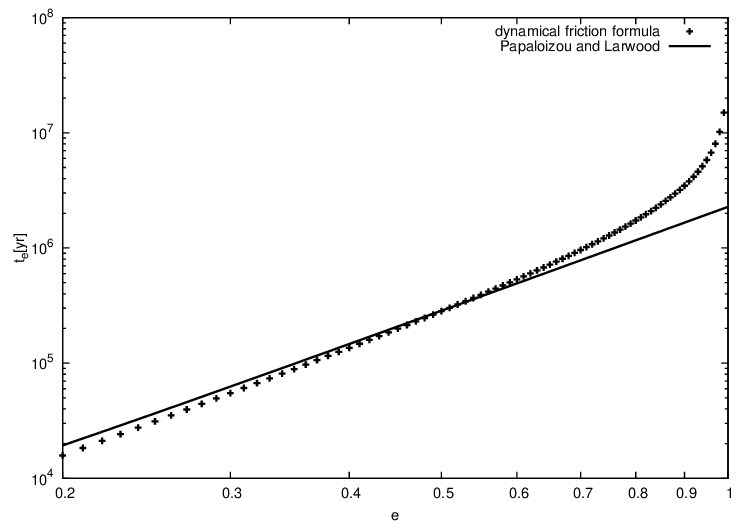}
 \caption{The migration timescale $t_m$ (left) and the eccentricity
 damping timescale $t_e$ (right) for fiducial model.  
 For both of the panels, we
 compare our results (crosses) and the formula by \citet{PL00} (solid
 line).  For $t_m$, we also show the formula of \citet{PL00} multiplied
 by $3/2$ is shown by dashed line for $e<0.5$, 
 which \citet{CN06} claims to fit the results of numerical simulations
 better. 
 }
 \label{fig:tmtecomp}
\end{figure}


\begin{figure}
 \plottwo{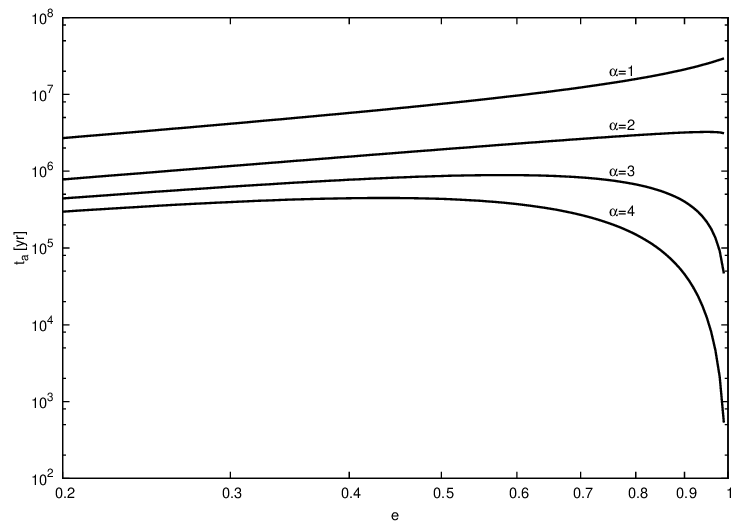}{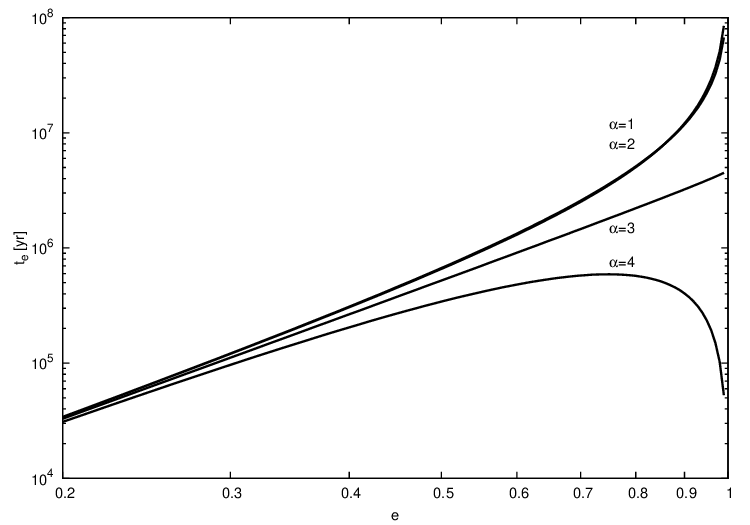}
 \caption{
 The timescale of the evolution of semimajor axis $t_a$ (left) and the 
 timescale of the evolution of eccentricity $t_e$ (right) for the models 
 with $\alpha=1,2,3,4$ ($p=0,1,2,3$, respectively for $q=1$).  
 The lines for $\alpha=1$ and $\alpha=2$ are almost identical for
 $t_e$.  We take $a/r_0=1$ and the unit of the time is year with
 $r_0=1\mathrm{AU}$ and $M_{\ast}=M_{\odot}$.  The direction of the
 semimajor axis evolution is inward, and the eccentricity always damps.   
 }
 \label{fig:tate_varalpha}
\end{figure}


\begin{figure}
 \plottwo{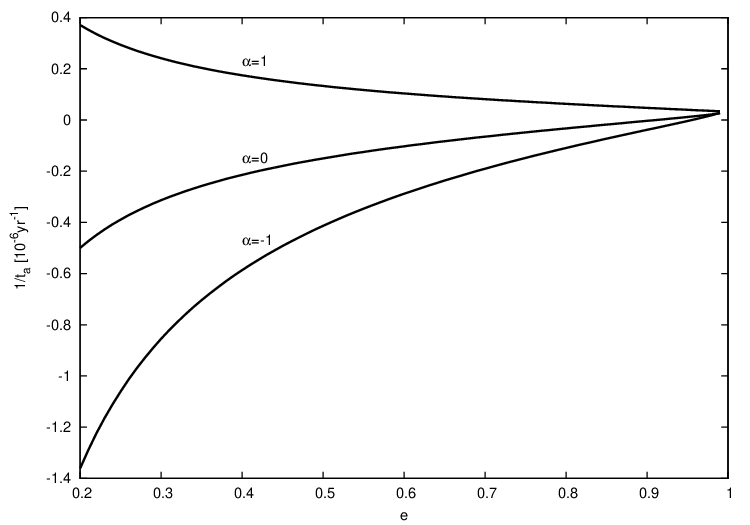}{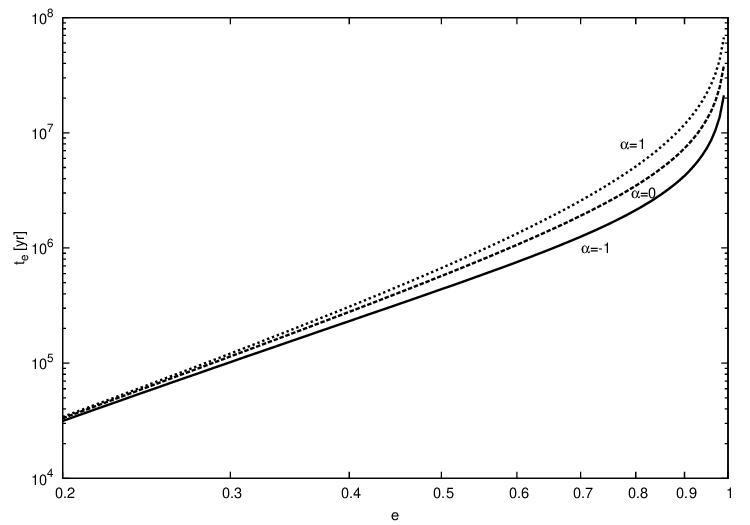}
 \caption{
 The timescale of the evolution of semimajor axis $t_a$ (left) and the
 timescale of the evolution of eccentricity $t_e$ (right) for the models
 with $\alpha=-1,0,1$ ($p=-2,-1,0$, respectively for $q=1$).  We take
 $a/r_0=1$ and the unit of the time is year with $r_0=1\mathrm{AU}$ and
 $M_{\ast}=M_{\odot}$.  The left panel shows $1/t_a$ for clarity, 
 since the sign of the direction of the semimajor axis evolution
 changes.   The negative sign indicates that the direction of the
 semimajor axis evolution is outward.   Eccentricity always decrease for
 the parameters presented here. 
 }
 \label{fig:tate_varalpha2}
\end{figure}


\begin{figure}
 \plottwo{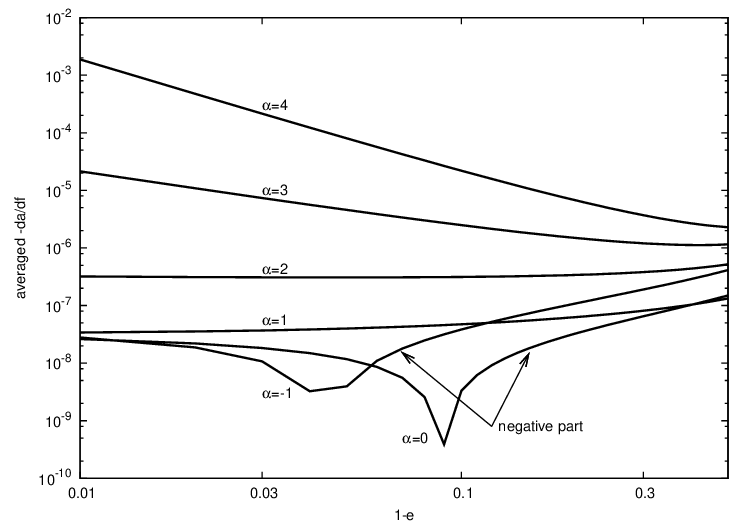}{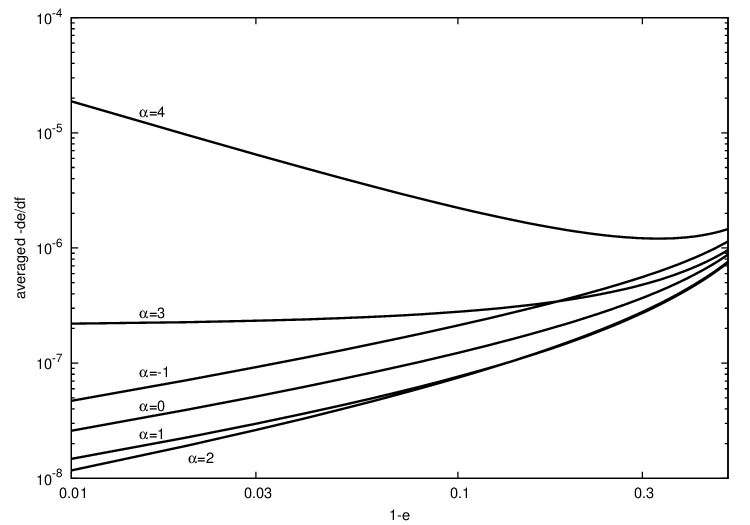}
 \caption{The orbit-averaged values of 
 $-da/df$ (left) and $-de/df$ (right) for various
 eccentricity and disk parameters with $a/r_0=1$.  
 The power $p$ (the parameter $\alpha=p+1$ in the case of $q=1$) 
 of the surface density
 is varied and denoted in the figures.  The horizontal axis shows the
 values of $1-e$.  For $da/df$, the absolute values are plotted if
 $-da/df$ is negative.  Only
 the cases when $\alpha=-1$ and $\alpha=0$ 
 with large values of $1-e$ 
 ($1-e>0.09$ for $\alpha=0$ and $1-e>0.05$ for $\alpha=-1$),
 we obtain negative values of $-\overline{da/df}$.  For the values of
 $-de/df$, the values are always positive.  
 }
 \label{fig:dade_ave}
\end{figure}

\clearpage

 \begin{deluxetable}{cccccccccc}
 \tabletypesize{\scriptsize}
 \tablecaption{Fitting results for $t_a$ \label{table:fitta}}
 \tablewidth{0pt}
 \tablehead{ \colhead{$\alpha$} & 
 \colhead{$C_{a0}$[yr]} & \colhead{$C_{ap}$} & 
 \colhead{$\zeta_{a0}$} & \colhead{$\zeta_{ap}$} & 
 \colhead{$\lambda_{a0}$} & \colhead{$\lambda_{ap}$} & 
 \colhead{$\eta_{a0}$} & \colhead{$\eta_{ap}$} & 
 \colhead{max. error} }
 \startdata
  1.0  &  3.789E+06  & -5.241E-01  &  9.759E-01  & -1.471E-05  &  1.320E-03  & -1.211E-01  & -1.986E-01  &  3.019E-09  &  1.886E-01 \\
  1.5  &  3.762E+06  &  4.353E-07  &  8.977E-01  &  3.777E-08  &  1.658E-01  &  2.010E-06  & -2.173E-01  &  1.068E-09  &  3.895E-02 \\
  2.0  &  3.988E+06  &  5.000E-01  &  1.008E+00  & -1.188E-09  &  1.129E+00  &  7.216E-08  &  5.318E-02  &  3.436E-09  &  4.497E-03 \\
  2.5  &  2.861E+06  &  1.000E+00  &  9.952E-01  &  0.000E+00  &  2.112E+00  &  0.000E+00  &  4.984E-01  &  0.000E+00  &  2.038E-03 \\
  3.0  &  2.265E+06  &  1.500E+00  &  9.944E-01  &  0.000E+00  &  2.063E+00  &  0.000E+00  &  9.957E-01  &  0.000E+00  &  2.787E-03 \\
  3.5  &  1.948E+06  &  2.000E+00  &  1.010E+00  &  1.604E-10  &  1.920E+00  & -3.424E-10  &  1.503E+00  & -4.489E-11  &  3.085E-03 \\
  4.0  &  1.793E+06  &  2.500E+00  &  1.041E+00  &  6.360E-10  &  1.774E+00  & -1.094E-09  &  2.012E+00  & -1.386E-10  &  1.058E-02 \\
 \enddata
 \end{deluxetable}


 \begin{deluxetable}{cccccccccc}
 \tabletypesize{\scriptsize}
 \tablecaption{Fitting results for $t_e$ \label{table:fitte}}
 \tablewidth{0pt}
 \tablehead{ \colhead{$\alpha$} & 
 \colhead{$C_{e0}$[yr]} & \colhead{$C_{ep}$} & 
 \colhead{$\zeta_{e0}$} & \colhead{$\zeta_{ep}$} & 
 \colhead{$\lambda_{e0}$} & \colhead{$\lambda_{ep}$} & 
 \colhead{$\eta_{e0}$} & \colhead{$\eta_{ep}$} & 
 \colhead{max. error} }
 \startdata
  1.0  &  3.426E+04  & -5.090E-01  &  2.651E+00  & -1.908E-06  &  1.158E-03  & -1.331E-02  & -6.765E-01  &  8.473E-10  &  8.247E-02 \\
  1.5  &  3.768E+04  & -9.810E-02  &  2.615E+00  & -3.975E-05  &  2.431E-03  & -1.324E-01  & -7.394E-01  & -2.246E-08  &  1.126E-01 \\
  2.0  &  3.307E+06  &  5.000E-01  &  2.914E+00  &  0.000E+00  &  1.258E+00  & -9.885E-11  & -7.597E-01  &  0.000E+00  &  5.556E-02 \\
  2.5  &  3.956E+06  &  1.000E+00  &  2.984E+00  &  0.000E+00  &  1.733E+00  & -4.559E-11  & -5.236E-01  &  0.000E+00  &  9.257E-03 \\
  3.0  &  2.372E+06  &  1.500E+00  &  2.991E+00  &  0.000E+00  & -7.393E-09  & -5.798E-09  & -2.964E-02  &  0.000E+00  &  1.802E-02 \\
  3.5  &  4.528E+06  &  2.000E+00  &  3.038E+00  &  0.000E+00  &  1.376E+00  & -5.197E-10  &  5.102E-01  &  0.000E+00  &  9.239E-03 \\
  4.0  &  7.059E+06  &  2.500E+00  &  3.183E+00  &  0.000E+00  &  8.549E-01  & -5.166E-10  &  1.027E+00  &  0.000E+00  &  2.573E-02 \\
 \enddata
 \end{deluxetable}

\end{document}